\documentclass[reprint,pre,aps,epsf,twocolumn,superscriptaddress]{revtex4}

\usepackage{graphicx}
\usepackage{amssymb}
\usepackage{amsmath}
\usepackage{tikz}
\usetikzlibrary{arrows,shapes,trees}

\def\3{\ss }           % scharf-ss
\begin{document}
\title{Kinetic theory for systems of self-propelled particles with metric-free interactions}
\author{Yen-Liang Chou}
\affiliation{Department of Physics, North Dakota State University,
Fargo, North Dakota, 58108-6050}
\author{Rylan Wolfe}
\affiliation{Department of Physics, North Dakota State University,
Fargo, North Dakota, 58108-6050}
\affiliation{Department of Chemistry and Biochemistry, North Dakota State University, 
Fargo, North Dakota 58105-5516}
\author{Thomas Ihle}
\affiliation{Department of Physics, North Dakota State University,
Fargo, North Dakota, 58108-6050}
\affiliation{Max Planck Institute for the Physics of Complex Systems, N{\"o}thnitzer Stra{\ss}e 38, 01187 Dresden, Germany}

\begin{abstract}
A model of self-driven particles similar to the Vicsek model [Phys. Rev. Lett. 75 (1995) 1226] 
but with metric-free interactions is
studied by means of a novel Enskog-type kinetic theory. 
In this model, N particles of constant speed $v_0$ try to align their travel directions with the average direction of a fixed number
of closest neighbors. 
At strong alignment a global flocking state forms. The alignment is defined by a stochastic rule, not by a Hamiltonian. 
The corresponding interactions are of genuine multi-body nature. 
The theory is based on a Master equation in 3N-dimensional phase space, which is made tractable by means of the molecular chaos approximation.
The phase diagram for the transition to collective motion
is calculated and compared to direct 
numerical simulations. 
A linear stability analysis of a homogeneous ordered state is performed using the kinetic but not the 
hydrodynamic equations in order to achieve high accuracy.
In contrast to the regular metric Vicsek-model 
no instabilities occur. This confirms previous direct simulations that for Vicsek-like models with metric-free interactions, 
there is no formation of 
density bands and that the flocking
transition is continuous.
\end{abstract}

\pacs{87.10.-e,05.20.Dd,64.60.Cn,02.70.Ns} 

\maketitle

{\small PACS numbers:87.10.-e, 05.20.Dd, 64.60.Cn, 02.70.Ns} 

\section{Introduction}
One of the most important unsolved problems in statistical physics is finding a global description of
far-from-equilibrium systems with many interacting objects.
Such a unified theory would be especially useful for
biological systems since they operate far from thermal equilibrium. 
Instead of looking for such a general theory we
focus on a minimal nonequilibrium model which
still displays interesting physics such as pattern formation and collective motion.
The goal is to provide inspiration for a more general approach by  
constructing a quantitative theoretical framework for the minimal model.
We consider a  model similar to the Vicsek-model (VM) of self-propelled particles \cite{vicsek_10,ramas_10}
which is simple enough to be treated numerically and
analytically. 
The VM was introduced in 1995 to describe the swarming of fish and birds \cite{vicsek_95_07}.
In this model, pointlike particles are driven with a
constant speed. At each time step, a given particle assumes the average
direction of motion of the particles in its neighborhood, %
with some added noise.
This model constitutes a dynamical version of the 2D XY model, because the velocity of the ``bird'', like the spin of the XY model, also has fixed magnitude and continuous rotational symmetry.
As the amplitude of the noise decreases, the system undergoes a phase transition from a
disordered state, in which the particles have no preferred global direction, to an
ordered state, in which
the particles move collectively in the same direction.
Hence, unlike the XY model, Vicsek's model exhibits long-range orientational order at non-zero noise.
This surprising fact
motivated renormalization group studies
by Toner and Tu \cite{toner_95_98} which confirmed the 
stabilization of the ordered phase far from the flocking threshold.
The phase transition was
originally thought to be continuous \cite{vicsek_95_07}, but recent numerical
work \cite{chate_04_08} indicates that the transition is discontinuous with
strong finite-size effects. 
The numerical studies also revealed that large density waves develop right next to the threshold while still maintaining
global orientational order.

Recently, it was shown by means of an Enskog-like kinetic theory that
the ordered phase of the VM is linearly unstable near the threshold \cite{ihle_11}. This instability was proposed as a possible explanation
for the density waves and discontinuous nature of the phase transition.
Another study for a related model with continuous time 
found a similar instability
%came to similar qualitative conclusions 
by means
of a Boltzmann equation \cite{bertin_06_09}, see also Ref. \cite{mishra_10}.
The VM assumes interaction with all neighbors 
within a fixed metric distance.
Recent experiments by Ballerini {\em et al} \cite{ballerini_08} on flocks of several thousand 
starlings indicate that this interaction rule might not be appropriate
for animal flocks.
Instead, it was discovered that each bird interacts on average with a fixed number of neighbors, typically six to seven.
This constitutes a topological or metric-free interaction because not the metric distance is relevant 
but who are the closest
neighbors. 
Ballerini {\em et al} argue further that due to evolutionary pressure the main goal of interaction among individuals is to maintain cohesion.
By comparing simulations with the regular VM and a modified VM with metric-free interactions they found that flocks, when facing predators,
kept cohesion
much better in the metric-free model.
This further supports the idea that
metric-free interactions should be dominant in animal flocks. 
While quite a number of analytical and numerical studies about self-propelled particles with metric interactions have been published 
\cite{aldana_03,huepe_04,aldana_07,peruani_06,peruani_10,peruani_08,menzel_12,roman_12,gao_11}, 
not much exists
for topological interactions \cite{chate_10,niizato_11,peshkov_12}.
In particular, no rigorous theory for the metric-free model of Ref. \cite{ballerini_08} exists.
In order to construct a theory which can be applied directly to this minimal computer model 
as well as to
experiments we adopted the original genuine multi-particle interactions and did not restrict 
ourselves to 
binary interactions.
Since other 2D experiments on shoaling fish estimate the number of tracked neighbors to be between three to five \cite{tegeder_95}
we explored a range of interaction partner numbers between two and seven.

The main results of this paper are a) the rigorous derivation of an Enskog-like 
kinetic equation for the one-particle density
for the metric-free model of Ref. \cite{ballerini_08} from first principles, and b) 
a linear-stability analysis of this kinetic equation, which showed that the flocking state is linearly
stable against perturbations of any wavelength.
%A brief discussion of these results was already presented in \cite{chou_talk_12}.

The remainder of this paper is structured as follows. In Sec. II we define the metric-free model.
In Sec. III we set up an exact equation for the $N$-particle probability density and derive 
the kinetic equation Eq. (\ref{COLLISION4}). In Sec. IV homogeneous solutions of this equation are discussed and the phase diagram of
the order-disorder transition is calculated. Sec. V deals with the linear stability analysis of the ordered phase, 
and Sec. VI describes direct simulations. A summary is given in Section VII.
Details concerning hypergeometric functions, exact solutions for special cases and integral tables are relegated to Appendix A, B, and C, respectively.
In Appendix D the Enskog kinetic approach is compared with the corresponding Boltzmann approximation.

\section{Model}

We consider a metric-free version of the VM, which was introduced in Ref. \cite{ballerini_08}.
This two-dimensional model consists of $N$ pointlike particles with continuous 
spatial coordinates ${\bf r}_i(t)$ and velocities ${\bf v}_i(t)$ which evolve via two
steps:
streaming and collision. 
During a time step $\tau$, particles stream ballistically:
${\bf x}_i(t+\tau)={\bf x}_i(t)+\tau {\bf v}_i(t)$.
The magnitude of the particle velocities is fixed to $v_0$. 
Only the directions $\theta_i$ of 
the velocity vectors 
are updated in the collision step by first finding the $M-1$ {\em closest} neighbors for a given particle $i$
where $M\ge 2$ is a fixed parameter. 
The directions of motion of particle $i$ and its neighbors 
determine the average direction,
\begin{equation}
\label{INTERACTION}
\Phi_i={\rm arctan}\left[{\sum_j^M {\rm sin}(\theta_j) \over \sum_j^M {\rm cos}(\theta_j)}\right]\,. 
\end{equation}
The new flying directions 
follow as
$\theta_i(t+\tau)=\Phi_i(t)+\xi_i$, where $\xi_i$ is a random number
chosen with a uniform probability from the interval $[-\eta/2,\eta/2]$.
In the context of the VM \cite{vicsek_10,ramas_10} this constitutes an angular noise model
with noise strength $\eta$.
The particles are always updated in parallel.
Note, that in the original VM \cite{vicsek_95_07} the number of interaction partners is fluctuating and density-dependent,
whereas the range of interactions is fixed to the radius $R$ of a circle around a given particle.
It is the opposite in the metric-free model. Here, the number of interaction partners, $M$, 
is a fixed parameter
but the interaction range fluctuates.
For example, for small local density $\rho({\bf r})$, the next neighbors can be far away but
no matter how geographically isolated a particle is, it is always connected with $M-1$ others via
the metric-free interaction rule, Eq. (\ref{INTERACTION}). 
To potentially allow comparison with  
experiments \cite{ballerini_08,cavagna_10} and simulations \cite{chate_10}, 
a large number $M=5$ to $8$ can be chosen.

We define an effective interaction range $R_{\text{eff}}$
by integrating the density over a circle 
and equating the result with the partner number $M$:
\begin{equation}
M=\int_{\circ} \rho({\bf r})\,d{\bf r}=\pi R_{\text{eff}}^2 \rho_0
\end{equation}
resulting in 
\begin{equation}
\label{REFF}
R_{\text{eff}}=\sqrt{M\over \pi \rho_0}
\end{equation}
Another important length scale is the mean free path (mfp)
given by the distance a particle travels between collisions,
\begin{equation}
\label{MFP}
\lambda=\tau\,v_0\,.
\end{equation}
Note that the mfp is {\em density-independent} in VM-like models because of the discrete nature of
the dynamics and because the particles have zero volume. 

The metric-free model is then characterized by four dimensionless control parameters:
the noise strength $\eta$, the ratio of the mfp to the effective interaction radius, $\Lambda=\lambda/R_{\text{eff}}$,
the partner number $M$, and the normalized system size, $\tilde{L}=L/\lambda$, of the $L\times L$ simulation box with periodic
boundary conditions. Secondary parameters of less obvious physical relevance, such as 
the total particle number $N$ and the average density $\rho_0=N/L^2$ can be easily expressed in terms of the four 
characteristic parameters, for example, using Eq. (\ref{REFF}) one finds, 
$N=M\Lambda^2 \tilde{L}^2/\pi$. The thermodynamic limit, $N\rightarrow \infty$, is 
then equivalent to $\tilde{L}\rightarrow \infty$ while keeping $M$, $\eta$ and $\Lambda$ constant.

\section{Kinetic theory}

Recently, a kinetic formalism for the Vicsek model beyond the Boltzmann theory has been developed \cite{ihle_11}.
Such an approach is particularly useful
for the metric-free VM, where particles always interact
with about four to seven neighboring particles at once. These genuine multi-body interactions cannot be described
by the binary collision approximation of the Boltzmann equation \cite{peshkov_12}.

The starting point for the kinetic formalism 
is a discrete-time Master equation in 3N-dimensional phase space 
for the N-particle probability density 
\begin{eqnarray}
\nonumber
& &P(\theta^{(N)}, {\bf X}^{(N)}+\tau {\bf V}^{(N)},t+\tau)= 
{1\over \eta^N}
\int_{-\eta/2}^{\eta/2}
d\xi^{(N)} \\
& & \times \int_0^{2\pi} d\tilde{\theta}^{(N)}
\,P(\tilde{\theta}^{(N)}, {\bf X}^{(N)},t) 
\label{PRL1}
\prod_{i=1}^N \hat{\delta}(\theta_i-\xi_i-\Phi_i) 
\end{eqnarray}
where 
${\bf X}^{(N)}\equiv({\bf x}_1,{\bf x}_2,\ldots, {\bf x}_N)$ and
$\theta^{(N)}\equiv(\theta_1,\theta_2,\ldots, \theta_N)$. 
The periodically continued delta function 
$\hat{\delta}(x)=\sum_{m=-\infty}^{\infty}\delta(x+2\pi m)$
accounts for angular periodicity, $\theta\equiv \theta+2\pi m$.
The velocities
${\bf V}^{(N)}\equiv({\bf v}_1,{\bf v}_2,..., {\bf v}_N)$, are given in terms
of angle variables, ${\bf v}_i=v_0(\cos{\theta_i}, \sin{\theta_i})$.
The collision integral contains integrations over
the pre-collisional angles $\tilde{\theta}_j$
and over N independent sources of angular noise.
This equation is exact and can also be interpreted as the discrete-time analogue
of the 
Liouville equation. Equations of this type have been used before, for example,  to analyze particle-based simulation methods
for fluid flow \cite{malev_99,ihle_09}.

Assuming that particles are uncorrelated 
{\it prior} to collisions,
the probability distribution can be expressed as a product of identical one-particle 
probability distributions:
$P(\theta^{(N)}, {\bf X}^{(N)})
=\prod_{i=1}^N P_1(\theta_i, {\bf x}_i)$.
This approximation of {\em molecular chaos} (MC) 
is valid at moderate and large noise strength $\eta$ and large 
mean free path $\lambda=\tau\,v_0$ compared to the effective interaction radius $R_{\text{eff}}$.
It can be seen 
as a dynamic mean-field approximation
because it neglects pre-collisional correlations.

The assumption of large mean free path, $\Lambda=\lambda/R_{eff}\gg 1$ is not very realistic for a system of swarming 
agents
because it would allow agents to bypass others at very close distances without including them in the subsequent 
interaction. Nevertheless, the MC ansatz is very useful for several reasons. First, approximations sometimes turn out to have a much larger 
range of validity than expected. 
For example, in a particle-based simulation method for fluid flow \cite{ihle_05}, the MC approximation gave correct results for most 
transport coefficients
down to $\Lambda=0.1$.
Second,  
the MC ansatz generates the lowest order terms in an expansion of the exact kinetic theory in the parameter $\varepsilon=1/\Lambda$ 
and thus can be seen as the first step towards a more complete theory.
In this paper, we analyze these MC contributions and leave
higher order terms for future work.
%. Work on higher order terms is in progress.

The usual procedure \cite{malev_99,ihle_09} to derive a kinetic equation for
the one-particle distribution function,
$f(\theta,{\bf x},t)=NP_1(\theta,{\bf x},t)$, is to multiply
the N-particle equation with the microscopic one-particle density,
$\sum_i\delta(\theta-\theta_i)\delta({\bf x}-{\bf x}_i)$, 
for the field variables $(\theta,{\bf x})$, 
and to integrate
over all phases, that is all particle positions $x_i$ and angles $\theta_i$. 
The left hand side of eq. (\ref{PRL1}) reduces then to
$NP_1(\theta,{\bf x}+\tau\,{\bf v},t+\tau)=f(\theta,{\bf x}+\tau\,{\bf v},t+\tau)$
because integrating out $k$ phases leads to $N-k$ particle probabilities such as in
the following example,
\begin{equation}
\int d\theta_i\,d{\bf x}_i P(\theta^{(N)}, {\bf X}^{(N)})
=P(\theta^{(N-1)}, {\bf X}^{(N-1)})\,.
\end{equation}

The collision term on the r.h.s. of Eq. (\ref{PRL1}) becomes 
\begin{eqnarray}
\nonumber
&I(\theta,{\bf x})&={N\over \eta}\int_{-\eta/2}^{\eta/2}\,d\xi\,\int_V d{\bf x}_2\ldots d{\bf x}_N\,
\int_0^{2\pi}\,d\tilde{\theta}_1\,d\tilde{\theta}_2\ldots d\tilde{\theta}_N
\\
\label{COLLISION1}
&\times&\hat{\delta}(\theta-\xi-\Phi_1)
{f(\tilde{\theta}_1,{\bf x})\over N}
{f(\tilde{\theta}_2,{\bf x}_2)\over N}\ldots
{f(\tilde{\theta}_N,{\bf x}_N)\over N}\,.
\end{eqnarray}
The overall prefactor $N$ results from the fact that the particles are physically identical and thus every $\delta(\theta-\theta_i)\delta({\bf x}-{\bf x}_i)$ term in the one-particle density yields the same contribution.
In Eq. (\ref{COLLISION1}), without loss of generality, particle $1$ was chosen to play a preferred role. 
Its position ${\bf x}_1$ is fixed to the field point 
${\bf x}$ but the positions of the other particles $2,3,\ldots N$ must be integrated over.
These integrations are more difficult than they appear because the average angle $\Phi_i$ has an implicit
dependence on all particle positions. For example, in a system with $M=4$ interaction partners, if particles 2,5, and 7 happen to be 
the closest ones to particle 1, only they are included in the calculation of the average angle, 
$\Phi_1=\Phi_1(\tilde{\theta}_1,\tilde{\theta}_2,\tilde{\theta}_5,\tilde{\theta}_{7})$,
but none of the others.
If for example, ${\bf x}_2$ is moved further away from ${\bf x}$, another particle, say number 9, 
might become one of the closest four  
and will replace particle 2 in the calculation, thus 
$\Phi_1=\Phi_1(\tilde{\theta}_1,\tilde{\theta}_9,\tilde{\theta}_5,\tilde{\theta}_7)$.
For yet another spatial arrangement of particles, $\Phi_i$ is determined by a different set of $M$ particles. 
That means, the $2N-1$ position and angular integrals are not independent of each other, they are coupled by the singular collision
kernel $\hat{\delta}(\theta-\xi-\Phi_1)$.
Fortunately, it is possible to rearrange the collision term and to reduce the number of integrals from infinity
(in the thermodynamic limit, $N\rightarrow \infty$) to the small finite number $2M-1$. 
In contrast to other kinetic approaches \cite{peshkov_12} no additional 
approximations are required (see also Appendix D). 

\begin{figure}
\begin{center}
\vspace{0cm}
\includegraphics[width=1.8in,angle=0]{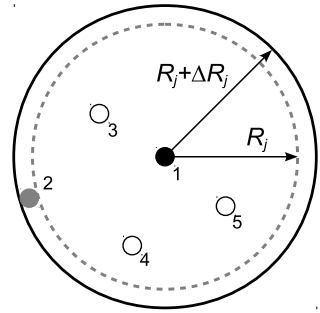}
\vspace*{-2.0ex}
\caption{
Illustration for the derivation of Eq. (\ref{COLLISION2}) with $M=5$ collision partners.
The selected particle $1$ is fixed to the center of the circle of radius $R_j$.
Particle 2 is integrated over the ring with inner radius $R_j$ and outer radius $R_{j+1}=R_j+\Delta R$ whereas
the remaining $M-2$ collision partners are integrated over the inner circle.
}
\label{FIG1}
\end{center}
\vspace*{-3ex}
\end{figure}

The main idea is to draw a set of concentric circles with radii $R_j=j\,\Delta R$, $j=1,2\ldots \infty$, around
particle $1$ at ${\bf x}_1={\bf x}$. The circle distance, $\Delta R$, will eventually become infinitesimal.
Next, one picks $M-1$ particles out of the available $N$ particles, puts {\em one} of them in the ring $j$ (between $R_{j+1}$ and $R_j$) and distributes
the other $M-2$ ones inside the circle of radius $R_j$,
see Fig. \ref{FIG1}.
The remaining $N-M$ particles are placed outside the ring $R_{j+1}$. There is $(N-1)(N-2)!/((M-2)! (N-M)!)$ possibilities
to segregate particles into three general areas: outside, inside the inner circle and inside the ring.
The outside particles are allowed to move over the entire space but without crossing into the ring or the inner circle. 
The single particle in the ring can move inside the ring only,
and the inner circle particles are allowed to take any position within that circle but cannot cross into
other areas.
For fixed ring label $j$ these rules describe all possibilities to have exactly $M$ interaction partners 
within a circle of radius $R_{j+1}$ but not within a smaller circle $R_j$. 
Situations where there is more than one particle in the ring are irrelevant
in the limit $\Delta R\rightarrow 0$ because the probability for these events goes faster to zero than the one for single occupancy of the ring.
By increasing $j$ to $j+1$, that is by going to the next larger ring and redistributing particles into the three zones, one realizes that none of the new configurations could have already occurred at smaller $j$.
We have thus constructed an alternative description of the particle positions 
in terms of ring number $j$, particle labels and positions inside three distinct zones.
For $\Delta R\rightarrow 0$ this representation is completely equivalent to the original description in terms of 
$({\bf x}_1,{\bf x}_2,\ldots, {\bf x}_N)$, meaning that no double counting or omission of configurations occur.
This new description is crucial for the evaluation of the position integrals
because for every configuration the identity of the $M$ interacting particles is fixed. 
The collision integral, eq. (\ref{COLLISION1}), can now be split in an infinite sum over the radii $R_j$ where
$M$ particles are inside $R_j+\Delta R$ and where $N-M$ particles are outside,
\begin{eqnarray}
\nonumber
&I(\theta,{\bf x})&={N(N-1)(N-2)!\over \eta (M-2)!(N-M)!}
\int_{-\eta/2}^{\eta/2}\,d\xi\,
\int_0^{2\pi}\,d\tilde{\theta}_1\ldots d\tilde{\theta}_N \\
&\sum_{j=1}^{\infty}& 
\nonumber
\int_{R(j)}d{\bf x}_2
\int_{C(j)}d{\bf x}_3\ldots d{\bf x}_M
\int_{O(j)}d{\bf x}_{M+1}\ldots d{\bf x}_N
\\
\label{COLLISION2}
&\times &\hat{\delta}(\theta-\xi-\Phi_1)
{f(\tilde{\theta}_1,{\bf x})\over N}
{f(\tilde{\theta}_2,{\bf x}_2)\over N}\ldots
{f(\tilde{\theta}_N,{\bf x}_N)\over N}\,.
\end{eqnarray}
The index $R(j)$ at the first spatial integral denotes integration over a
thin ring centered around ${\bf x}$ with inner and outer radii $R_j=j\,\Delta R$ and $R_{j+1}=(j+1)\,\Delta R$, respectively.
The index $C(j)$ means that the integration goes over the entire interaction circle with radius $R_j$, and $O(j)$ denotes
integration over the area outside a circle with radius $R_{j+1}$.
The combinatorial prefactor can be simplified for large $N$ as 
\begin{equation}
{N!\over (M-2)!(N-M)!}\approx {N^M\over (M-2)!}
\end{equation}
and using the definition of the particle density $\rho$ as the zeroth moment of $f$,
\begin{equation}
\int_0^{2\pi}f(\theta,{\bf x})\,d\tilde{\theta}=\rho({\bf x}),
\end{equation}
the integration over the angles $\tilde{\theta}_{M+1}\ldots \tilde{\theta}_N$ 
of the outside particles can be performed,
\begin{eqnarray}
\nonumber
&I(\theta,{\bf x})&={1\over \eta (M-2)!}
\int_{-\eta/2}^{\eta/2}\,d\xi\,
\int_0^{2\pi}\,d\tilde{\theta}_1\ldots d\tilde{\theta}_M \\
&\sum_{j=1}^{\infty}& 
P_j({\bf x})
\int_{R(j)}d{\bf x}_2
\int_{C(j)}d{\bf x}_3\ldots d{\bf x}_M
\\
\label{COLLISION3}
& &\hat{\delta}(\theta-\xi-\Phi_1)
f(\tilde{\theta}_1,{\bf x})
\ldots
f(\tilde{\theta}_M,{\bf x}_M)
\end{eqnarray}
where $P_j$ is the contribution from the outside particles, % follows as %could be separated as
\begin{equation} 
\label{OUTSIDE1}
P_j({\bf x})=
\int_{O(j)}d{\bf x}_{M+1}\ldots d{\bf x}_N
{\rho({\bf x}_{M+1})\over N}
{\rho({\bf x}_{M+2})\over N}\ldots
{\rho({\bf x}_{N})\over N}\,.
\end{equation} 
Defining the average particle number in a circle of radius $R_{j+1}$ centered around ${\bf x}$ as
\begin{equation} 
\label{LOCAL_M}
\overline{M}_j({\bf x})=\int_{C(j+1)} \rho({\bf x}')\,d{\bf x}' 
\end{equation} 
and using the fact that integrating the density over the entire space is equal to the total particle number $N$,
\begin{equation} 
\label{ALLSPACE}
\int_\text{all space} \rho({\bf x}')\,d{\bf x}'=N 
\end{equation} 
one finds that
\begin{equation} 
\label{PJ1}
\int_{O(j)} {\rho({\bf x}')\over N}\,d{\bf x}'=1-{\overline{M}_j({\bf x})\over N} 
\end{equation} 
which, in the thermodynamic limit, gives
\begin{equation} 
\label{PJ2}
P_j=\lim_{N\rightarrow \infty} \left(1-{\overline{M}_j\over N}\right)^{N-M}={\rm e}^{-\overline{M}_j} 
\end{equation} 
Combining Eqs. (\ref{COLLISION3}) and (\ref{PJ2}) 
leads to the final evolution equation for the one-particle density
of the metric-free model
\begin{eqnarray}
\nonumber
& &f(\theta,{\bf x}+\tau\,{\bf v}(\theta),t+\tau)= \\
\nonumber
& &\lim_{\Delta R\rightarrow 0}\, {1\over (M-2)!}
\int_{-\eta/2}^{\eta/2}\,{d\xi\over \eta}\,
\int_0^{2\pi}\,d\tilde{\theta}_1\ldots d\tilde{\theta}_M \\
\label{COLLISION4}
&\sum_{j=1}^{\infty}& 
{\rm e}^{-\overline{M}_j({\bf x})} 
\int_{R(j)}d{\bf x}_2
\int_{C(j)}d{\bf x}_3\ldots d{\bf x}_M
\\
\nonumber            
&\times &\hat{\delta}(\theta-\xi-\Phi_1)
f(\tilde{\theta}_1,{\bf x})
f(\tilde{\theta}_2,{\bf x}_2)
\ldots
f(\tilde{\theta}_M,{\bf x}_M)
\end{eqnarray}
Eq. (\ref{COLLISION4}) has a highly nonlocal and nonlinear collision term.
For example, the exponent $\overline{M}_j$ is a functional 
of the density $\rho$ which
is itself a functional of $f$, $\overline{M}_j=\overline{M}_j[\rho[f]]$. However, this equation is still analytically
tractable.
Note the integrations across interaction radii which describe collisional momentum transfer --
a key feature of the Enskog equation -- and absent in Boltzmann approaches \cite{bertin_06_09,peshkov_12}, see Appendix D.
\section{Homogeneous solutions and phase diagram}

It is useful to first study 
homogeneous stationary solutions,
$f(\theta,\mathbf{x},t)=\bar{f}(\theta)$,
of the Enskog-like kinetic equation
(\ref{COLLISION4}).
The integrands are now independent of position, and,
for infinitesimal $\Delta R$, the ring integral and the circle integrals in (\ref{COLLISION4}) can be replaced by
\begin{eqnarray}
\nonumber
\int_{R(j)}\,d{\bf x}_2&\simeq &2\pi R_j\,\Delta R \\
\nonumber
\int_{C(j)}\,d{\bf x}_3&=&\pi R_j^2\;{\rm and} \\
\overline{M}_j&=&\pi (R_j+\Delta R)^2\rho_0\simeq\pi R_j^2\rho_0\,.
\end{eqnarray}
For $\Delta R\rightarrow 0$, the sum $\sum_j\,\Delta R$ goes over to the integral $\int\,dR$, and after substituting
$z=R\sqrt{\pi \rho_0}$ one finds for the r.h.s. of Eq. (\ref{COLLISION4}),
\begin{eqnarray}
\nonumber
I(\theta)&=&{\Psi_M(\theta)\over \rho_0^{M-1}} {2\over (M-2)!}\int_0^{\infty}z^{2M-3}{\rm exp}(-z^2)\,dz \\
\nonumber
\Psi_M(\theta)&=&
\frac{1}{\eta}\int_{-\eta/2}^{\eta/2}d\xi\int d\tilde{\theta}_1\ldots d\tilde{\theta}_M
\delta(\theta-\xi-\Phi_1)\\
\label{PSI_DEF}
&&\times \bar{f}(\tilde{\theta}_1)  \bar{f}(\tilde{\theta}_2)\ldots  \bar{f}(\tilde{\theta}_M).
\end{eqnarray}
The integral over $z$ is solvable for all $M$,
\begin{equation}
\label{Z_INTEGRAL}
{2\over (M-2)!}\int_0^{\infty}z^{2M-3}{\rm exp}(-z^2)\,dz=1 
\end{equation}
and the kinetic equation (\ref{COLLISION4}) takes the form of a nonlocal
fixed point equation for the function $\bar{f}(\theta)$,
\begin{equation}
\label{HOMOGENFINAL}
\bar{f}(\theta)=I(\theta)={\Psi_M(\theta)\over \rho_0^{M-1}} 
\end{equation}
This equation constitutes a nonlinear singular Fredholm integral equation of the second kind.

\subsection{Disordered state}

From direct numerical simulations of VM-like models \cite{vicsek_95_07,chate_04_08} and from previous analytical work
on the regular VM \cite{ihle_11}, we expect $f=f_0=\rho_0/(2\pi)$
to be a fixed point of the collision integral, independent of noise strength and partner number $M$.
This constant solution describes the disordered phase of the model because it does not depend on the angle, and thus
every flying direction $\theta$ occurs with the same probability.
In order to check this expectation we expand the collision operator of Eq. (\ref{HOMOGENFINAL}) into an angular Fourier series
which regularizes the singular collision kernel
$\hat{\delta}(\theta-\xi-\Phi_1)$,
\begin{equation}
\label{FOURIERCOLL}
I(\theta)=C_0+\sum_{k=1}^\infty \left[C_k {\rm cos}(k\theta)+
H_k {\rm sin}(k\theta)\right]
\end{equation}
where 
\begin{eqnarray}
\nonumber
C_0&=&{1\over 2\pi} \int_0^{2\pi} I(\theta)d\theta\\
\nonumber
C_k&=&{1\over \pi} \int_0^{2\pi} {\rm cos}(k\theta)\,I(\theta)d\theta \\
\label{FOURIERCOLL}
H_k&=&{1\over \pi} \int_0^{2\pi} {\rm sin}(k\theta)\,I(\theta)d\theta
\;\;{\rm for}\;k>0
\end{eqnarray}
Since $f_0$ does not depend on $\theta$, only integrals of type 
$\int_0^{2\pi}{\rm cos}(k\Phi_1)\,d\tilde{\theta}_1\ldots d\tilde{\theta}_M$
and
$\int_0^{2\pi}{\rm sin}(k\Phi_1)\,d\tilde{\theta}_1\ldots d\tilde{\theta}_M$
occur
when $f_0=\rho_0/(2\pi)$ is inserted into $\Psi_M$ in Eq. (\ref{PSI_DEF}) and
$C_k$ and $H_k$ are evaluated.
These integrals vanish for nonzero $k$ because the average angle $\Phi_1$
takes all values
between $0$ and $2\pi$ with the same probability.
Hence, $C_k=H_k=0$ for $k>0$ and it
remains to find $C_0$.
After integrating over the noise and the pre-collisional angles 
one 
obtains $\int_0^{2\pi} \Psi_M\,d\theta=\rho_0^M$ and
\begin{equation}
C_0={\rho_0\over 2\pi}=f_0
\end{equation}
Thus, we indeed find that $f_0$ is always a fixed point of the collision operator $I$,
\begin{equation}
I[f_0]=f_0\,.
\end{equation}
This is a nice confirmation that the alternative representation
of the particle configuration in terms of rings, described in the previous section, is correct, and that
no relevant configurations were left out or overcounted.

\subsection{Ordered state}

An ordered state of self-propelled particles is characterized by particles
which have the same nonzero average flying direction.
Such a state 
breaks the rotational symmetry of the model 
and represents another fixed point of 
the integral equation (\ref{HOMOGENFINAL}).
Its one-particle density,
$f_{\text{ord}}$, depends on the angle and has a maximum at some arbitrary angle $\hat{\theta}$
which is the direction of ordered motion.
We choose $\hat{\theta}=0$  because then only Fourier cosine coefficients 
are needed,
\begin{equation}
\label{EXPAND1}
f_{\text{ord}}(\theta)=\sum_{k=0}^{\infty} g_k\,{\rm cos}(k\theta)
\end{equation}
The integral equation (\ref{HOMOGENFINAL}) reduces to an infinite set of algebraic equations, 
\begin{equation}
\label{FIXPOINT_FOURIER}
g_k=C_k(g_0,g_1,\ldots g_{\infty}) 
\end{equation}
where the $C_k$ are the Fourier coefficients of the collision operator $I(\theta)$, see Eq. (\ref{FOURIERCOLL}).
The calculations for $k=0$ are identical to the ones for the disordered phase analyzed above, thus
$g_0=C_0=\rho_0/(2 \pi)$.
To proceed, we assume that near a specific value of the noise, $\eta=\eta_C$,
only the lowest Fourier modes are relevant,
\begin{equation}
\label{ASYMP_G_K}
g_0\gg g_1\gg g_2\gg \ldots
\end{equation}
which can be easily verified {\em a posteriori}.
To find this critical noise $\eta_C$, all terms with $k>1$ are neglected, and only the equation for $k=1$
in (\ref{FIXPOINT_FOURIER}) is evaluated.
Inserting $f=\rho_0/(2\pi)+g_1{\rm cos}(\theta)$ into the collision operator and solving Eq. (\ref{FOURIERCOLL})
for $C_1$ yields 
\begin{eqnarray}
\nonumber
C_1&=&g_1 \Gamma+\mathcal{O}(g_1 g_2 g_0^{M-2}) \\
\label{OMEGA_KC1DEF}
\Gamma(\eta)&=&{4 M\over \eta}\sin{\eta\over 2}\,K_C^1(M) \\  
\nonumber            
K_C^1(M)&=& {1\over (2\pi)^M}\int_0^{2\pi} d\tilde{\theta}_1 \ldots d\tilde{\theta}_M
\cos{\Phi_1}
\cos{\tilde{\theta}_1}
\end{eqnarray}
Setting $C_1=g_1$, $\eta=\eta_C$ in Eq. (\ref{OMEGA_KC1DEF}) and using the asymptotic vanishing of all higher modes, Eq. (\ref{ASYMP_G_K}),
at the critical point
leads to an implicit equation for the critical noise, 
\begin{equation}
\label{DEFINE_CRIT}
\Gamma(\eta_C)=1\,. 
\end{equation}
The $M$-dimensional integral $K_C^1$ was calculated numerically for $2 \le M\le 20$ as well 
as analytically for $M=1,2,3$ and $M\rightarrow \infty$, see Appendix C and also Tables \ref{TAB1} and II. %\ref{TAB_M3_INT}.
These calculations are very similar to the ones for the network model of Ref. \cite{aldana_03}.
\begin{table}
\begin{center}
{\small
\begin{tabular}{ c  l l l l l l }
\hline
$M$             & $1$ & $2$    &  $3$    &  $4$     &  $5$     & $10$    \\ 
\hline 
$K_{c}^1$       & $1/2$ & $1/\pi$& 0.2624    & 0.2249     &  0.2008    & 0.141 \\                            
\hline
\end{tabular}
}
\caption{Analytical ($M=1,2,3$) and numerical results ($M\geq 4$)
for the integrals defined in Eq. (\ref{OMEGA_KC1DEF}).
The asymptotic behavior for $M\to \infty$ is, $K_c^1\sim \sqrt{\pi/(16 M)}$.
%$K_c^1\sim \sqrt{\dfrac{\pi}{16 M}}$. 
}
\label{TAB1}
\end{center}
\end{table}

\begin{figure}
\begin{center}
\vspace{0cm}
\includegraphics[width=3.2in,angle=0]{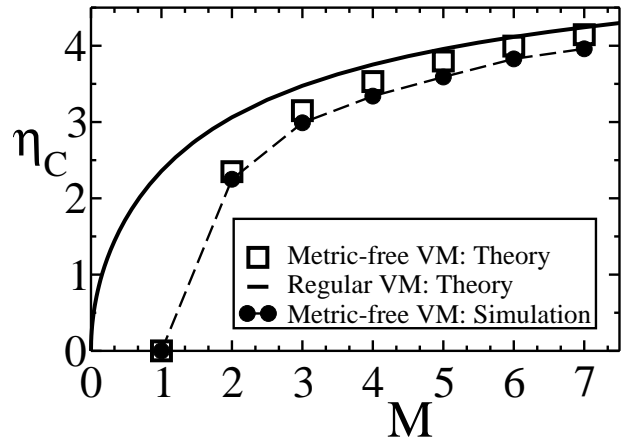}
\caption{The critical noise $\eta_C$ of the metric-free VM calculated from 
Eqs. (\ref{OMEGA_KC1DEF}, \ref{DEFINE_CRIT}) as a function of the 
number of collision partners $M$ 
in comparison with results for the regular VM from Ref. \cite{ihle_11}.
Direct simulation results for $N=5000$ particles and large $\Lambda=\lambda/R_{\text{eff}}=5.66$
are shown by the full circles.
If the noise $\eta$ is below the symbols, the system is found in the ordered phase with a non-zero total momentum.
The dashed line just serves as a guide to the eye.
}
\label{FIG2}
\end{center}
\vspace*{-5ex}
\end{figure}
The critical noise is plotted as function of partner number $M$ in Fig. \ref{FIG2}.
For large $M$, this mean field phase diagram agrees well with the 
one of the regular Vicsek model \cite{ihle_11}, and shows the same asymptotic behavior $\eta_C\rightarrow 2\pi$
for $M\rightarrow \infty$.
This is because for large $M$, the probability distribution for the actual radius of interaction becomes very narrow,
thus the observed radii of interaction are very close to the average radius $R_{\text{eff}}$.
Alternatively, this can be explained in the context of the regular VM. There, for large density,
the actual  particle number in a circle with fixed radius
is very close to the average number $M$. However, even 
at very large $M$,
when the critical noise is almost identical for both models,
there remains to be an important difference in the stability of the ordered phase
which we will analyze further below. As a result, the regular and the metric-free VM become only asymptotically identical, at infinite $M$.

\subsection{Order parameter calculations}

The physical meaning of $\Gamma$, defined in Eq. (\ref{OMEGA_KC1DEF}), can be understood by expressing the
average particle momentum ${\bf w}$ as the first moment of $f$,
\begin{eqnarray}
{\bf w}&=&\rho{\bf u}=\int_0^{2\pi}{\bf v}(\theta) f(\theta) d\theta\\
{\bf v}&=&v_0(\cos{\theta},\sin{\theta})\,,
\end{eqnarray}
where ${\bf u}$ is the local macroscopic velocity.
According to Eq. ({\ref{EXPAND1}) we also have 
\begin{equation}
\label{G1_W_LINK}
g_1={1\over \pi}\int_0^{2\pi} \cos{\theta}\,f(\theta)\,d\theta={w_x\over \pi v_0}
\end{equation}
Hence, the Fourier mode $g_1$ is proportional to the x-component of the momentum
whereas the y-component is zero in the special case of $\hat{\theta}=0$ considered here.
The quantity $\Gamma$ can be interpreted as the amplification factor of
momentum which can always be locally created or destroyed because the collision rules, Eq. (\ref{INTERACTION}), do not conserve momentum.
One finds that $\Gamma<1$ for $\eta>\eta_C$. Thus, above the critical noise, a small nonzero momentum
quickly goes to zero, and the system reaches the disordered phase with all $g_k=0$ for $k>0$.
Below the critical noise, $\eta<\eta_C$, the amplification factor $\Gamma$ is larger than unity, an initially small 
momentum is amplified, higher modes $g_2,\ldots g_{\infty}$ get excited until a stationary state with nonzero momentum is reached.
In order to quantitatively describe
this ordered state and to determine whether the order-disorder transition is continuous or discontinuous, 
the first few members of the hierarchy of equations have to be analyzed for $\eta<\eta_C$.
The smaller the noise $\eta$, the more members have to be included in order to achieve acceptable accuracy.
It is convenient to normalize the Fourier modes with $2 g_0=\rho_0/\pi$,
\begin{equation}
\label{NORMALG}
G_k={g_k\over 2 g_0} 
\end{equation}
because, for zero noise, all modes $g_k$, $k>0$ become equal to
$2 g_0$.
The normalized mode $G_1$ corresponds to the order parameter 
$|\sum_{i=1}^N{\bf v}_i|/(v_0 N)$ typically
used in direct simulations of flocking models \cite{vicsek_95_07,chate_04_08}.
In order to keep track of the relative sizes of terms, we introduce
the book keeping parameter $\epsilon$ and assume the scaling $G_k\sim\epsilon^k$.
This scaling was used previously \cite{ihle_11} and can be easily verified after the modes $G_k$
have been calculated, see Fig. \ref{FIG3}.
Including terms up to order $\epsilon^7$, the first six members of the fixed point equations in angular Fourier space, Eq. (\ref{FIXPOINT_FOURIER}),
for $M=2$ interaction partners were found as, 
\begin{eqnarray}
\nonumber
G_0&=&{1\over 2} \\
\nonumber
G_1&=&{A_1\over \pi}(2G_0 G_1-{1\over 3} G_1 G_2+{1\over 5} G_2 G_3-{1\over 7} G_3 G_4+\ldots \\
\nonumber
G_2&=&{A_2\over 4} G_1^2 \\
\nonumber
G_3&=&{A_3\over \pi}(G_1 G_2-{2\over 3} G_0 G_3+{1\over 5} G_1 G_4-{1\over 7} G_2 G_5+\ldots \\
\nonumber
G_4&=&{A_4\over 4} G_2^2 \\
\nonumber
G_5&=&{A_5\over \pi}(G_2 G_3-{1\over 3} G_1 G_4+{2\over 5} G_0 G_5-{1\over 7} G_1 G_6+\ldots\;\;{\rm with} \\
\label{G_M2}
A_k&=&{2^{M+1}\over \eta k}{\rm sin}\left({\eta k\over 2}\right)
\end{eqnarray}
The r.h.s of Eq. (\ref{G_M2}) contains only quadratic terms because
there is only binary interactions. For $M=3$ only cubic terms appear 
because all collisions are of three-body type.

The general structure of the fixed point equations for arbitrary $M$ and $k=1,2,\ldots\infty$ is given by
\begin{eqnarray}
\nonumber
& &G_k=A_k\sum_{\{i_j=0\}}^{\infty} J^{(k)}(i_1 i_2 i_3\ldots i_M)\, G_{i_1} G_{i_2}\ldots G_{i_M} \\
\nonumber      
& &J^{(k)}(i_1 i_2 i_3\ldots i_M)=\delta_k^{(M)} 
{1\over (2\pi)^M} \\
\label{GFORALL}
%\left(
%\prod_{n=1}^M
%\int_0^{2\pi}
%d\tilde{\theta}_n\,
%{\rm cos}(i_n\Phi_n)
%\right)
& &\times\int_0^{2\pi}
d\tilde{\theta}^{(M)}\,
{\rm cos}(k\Phi)
{\rm cos}(i_1\tilde{\theta}_1)
%{\rm cos}(i_2\tilde{\theta}_2)
\ldots
{\rm cos}(i_M\tilde{\theta}_M)
\end{eqnarray}
with $d\tilde{\theta}^{(M)}\equiv d\tilde{\theta}_1\,
d\tilde{\theta}_2\,
\ldots
d\tilde{\theta}_M$.
The Kronecker symbol 
\begin{equation}
\delta_k^{(M)}\equiv\delta_{k,\pm i_1\pm i_2\ldots i_M}\,
\end{equation}
emphasizes that an angular integral $J^{(k)}(i_1 i_2 i_3\ldots i_M)$ is nonzero only 
if some addition or subtraction of its lower indices is equal to the hierarchy level, 
$k=\pm i_1 \pm i_2\ldots i_M$.
For example, for $M=k=3$ the term $G_3 G_2 G_2$ would appear in the equation for $G_3$ since $3+2-2=3$
but not the term $G_1^2 G_2$ because $\pm 1 \pm 1 \pm 2$ is never equal to $3$ for any 
combination of plus and minus signs.
We expect this property to be a consequence of rotational symmetry but were not able to
find a mathematical proof.
These angular integrals are $M$-dimensional and can be evaluated analytically for 
certain cases such as $M=3$
but
it is easier to determine them numerically, see Appendix C.
However, even the numerical evaluation is very time consuming for high integral dimension $M\ge 7$ 
and for high mode numbers $k$ and $i_n$.

If the fixed point hierarchy, Eq. (\ref{GFORALL}) is truncated at level $k=k_T$ and only terms up to order
$\epsilon^{k_T}$ are included, a single algebraic equation of order $k_T-1$ can be derived for the order parameter $G_1$. 
This equation was solved for various truncation levels, see Fig. \ref{FIG3} a) for $M=2$, Fig. \ref{RYLAN_ORDER} for $M=3$, 
and Fig. \ref{RYLAN_DIFFERENTN} for $M=7$.                                     
Even for relatively large $k_T=5$, the accuracy quickly detoriates if the noise 
is smaller than about 50 \% of the critical noise, as seen in Fig. \ref{FIG3} a).
In order to overcome this restriction, 
the integral equation, Eq. (\ref{HOMOGENFINAL}), was solved directly 
by an iterative
numerical procedure which resolves the distribution function using 500 angular modes, 
details will be given elsewhere \cite{chou_12}.
In Fig. \ref{FIG3} one sees that this gives excellent accuracy even very close to zero noise
where $G_1=1$ is predicted analytically.
Once $G_1$ is known, all the higher modes can be calculated and serve as the ground state solution
in the stability analysis presented in the next section.

Evaluating the algebraic equation for $G_1$ near the critical noise gives
\begin{equation}
\label{CRIT_EXP}
G_1\sim\left({\eta_C-\eta\over \eta_C}\right)^{1/2}\,,
\end{equation}
thus, leading to the critical exponent $1/2$. This is expected for a mean field theory and was also found for the regular VM \cite{ihle_11}.
This means that at the mean-field level and as long as the ordered phase is stable against fluctuations near the flocking threshold, 
the order-disorder
transition is continuous.
One also finds the general scaling $G_k\sim [(\eta_C-\eta)/\eta_C]^{k/2}$, which is confirmed in Fig. \ref{FIG3} b),
Thus, the expansion parameter $\epsilon$ can be identified with
\begin{equation}
\label{DEF_EPS}
\epsilon\equiv\sqrt{\eta_C-\eta\over \eta_C}\,.
\end{equation}
\begin{figure}
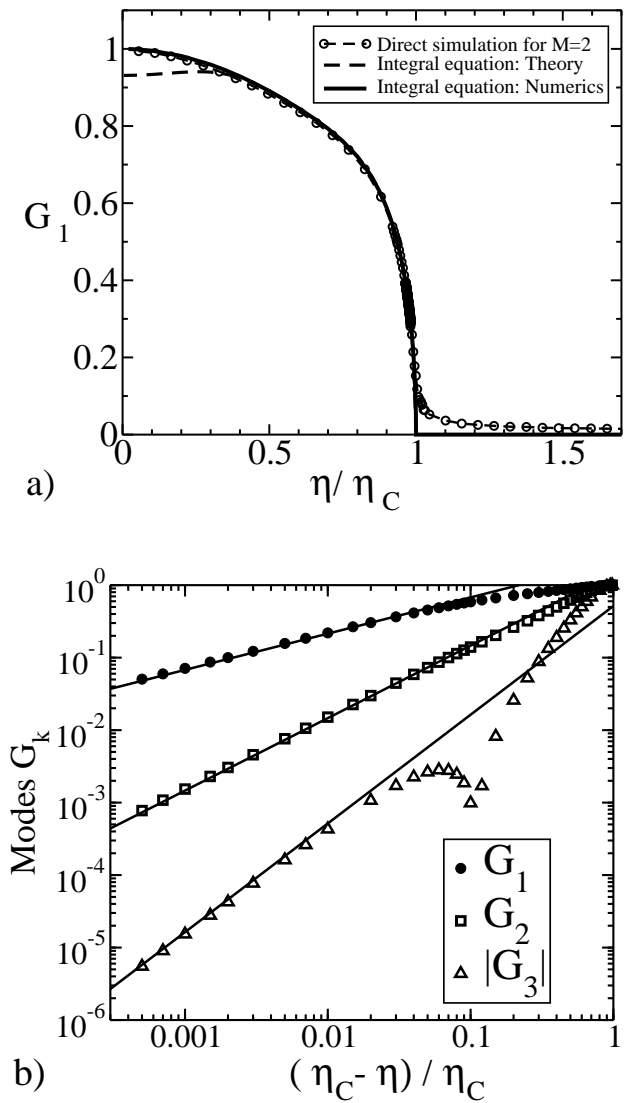

\begin{center}
\vspace{0cm}
\includegraphics[width=3.2in,angle=0]{FIG3a_CHOU.eps}

\vspace*{5ex}
\includegraphics[width=3.2in,angle=0]{FIG3b_CHOU.eps}
\vspace{-1ex}
\caption{
Order parameter $\Omega=G_1$ and higher modes as a function of the noise for partner number $M=2$.
a) Analytical calculations (dashed line) for $G_1$ from Eq. (\ref{G_M2}) are compared with the numerical solution (solid line)
of the integral equation (\ref{HOMOGENFINAL}) and with direct simulations (circles) of $N=5000$ particles at $\Lambda=5.66$.
The noise for the theoretical curves was rescaled by $\eta_{C,theo}=2.34923$, whereas the noise for the simulation curve
was rescaled by the slightly smaller critical noise $\eta_{C,sim}=2.2713$.
b) The first three modes $G_1$, $G_2$ and $G_3$ are numerically determined from Eq. (\ref{HOMOGENFINAL})
and plotted versus the distance to the threshold,
$\epsilon=(\eta_C-\eta)/\eta_C$ in order to verify the scaling $G_k\sim \epsilon^{k/2}$.
The straight lines correspond to the exponents $1/2$, $1$, and $3/2$, respectively .
Since $G_3$ is negative for $0.9\le \eta/\eta_C<1$ the absolute value of $G_3$ was plotted, which leads to the spurious dip
at $\epsilon\approx 0.1$.
}
\label{FIG3}
\end{center}
\vspace*{-5ex}
\end{figure}

\section{Linear stability analysis}

For the regular VM \cite{ihle_11} and related metric models \cite{bertin_06_09} it has been shown that 
the homogeneous ordered state 
is always
unstable to 
long wavelength perturbations in a small window $\eta_S\le \eta\le \eta_C$ 
right below the flocking threshold $\eta_C$.
The strongest instability occurs for longitudinal perturbations where the wave vector
${\bf k}$ of a perturbation is parallel to the average flying direction $\hat{n}$ of the homogeneous ground state.
Once the angle between $\hat{n}$ and ${\bf k}$ increases beyond a critical angle, the system is stable.
This linear instability explains the formation of large density bands in direct simulations of the VM \cite{vicsek_95_07,chate_04_08}.
These bands, however, have not been observed in simulations of the metric-free model \cite{ballerini_08,chate_10}.
This leads to the obvious hypothesis that the ordered state of the metric-free model is stable.
The question is whether it is linearly or nonlinearly stable or whether it is unstable but only at huge wavelengths
beyond the system sizes used in direct simulations.

In a previous paper \cite{ihle_11}, we had first derived the hydrodynamic equations from the kinetic description by a Chapman-Enskog procedure
and then analyzed the stability of the hydrodynamic equations.
Such a derivation is very tedious and involves several additional approximations such as considering 
only small spatial gradients and
assuming proximity to the flocking threshold where higher kinetic modes are enslaved to the lower ones.

Here, we employ a much faster and more accurate approach to the stability of the model.
Bypassing
the hydrodynamic description completely, we directly impose spatio-temporal perturbations into the kinetic
equation (\ref{COLLISION4}) and analyze their dynamics. Depending on the wavelength and 
the distance to the threshold, $\eta_C-\eta$,
the results of these calculations can be refined to the desired accuracy
by increasing the number of kinetic modes to be included.
The validity of hydrodynamic equations for only density and momentum 
is questionable anyway for 
models where momentum
is not conserved because 
momentum can be considered a slow variable only in special cases such as proximity to the threshold.
Further away from the threshold there is no a priori justification to neglect 
higher kinetic modes since their relaxation rates are not much different
from the one of the momentum.

Introducing a small perturbation 
\begin{eqnarray}
\nonumber
 \delta f(\theta,{\bf x},t)&=&\sum_{n=0}^{\infty}[ \delta g_n \cos{( n\theta)}+\delta h_n \sin{(n\theta})] \\
\delta g_n(\mathbf{x},t)&=&\delta \hat{g}_n~e^{i\mathbf{k}\cdot\mathbf{x}+\omega t}\nonumber\\
\delta h_n(\mathbf{x},t)&=&\delta \hat{h}_n~e^{i\mathbf{k}\cdot\mathbf{x}+\omega t},
\end{eqnarray}
of the homogeneous steady state
$\bar{f}(\theta)$, 
the distribution function changes to $f=\bar{f}+\delta f$.
The corresponding perturbations of the Fourier coefficients of $f$ are 
denoted as $\delta g_n$ and $\delta h_n$.
For brevity we will omit the time argument
in $f$, $\delta g_n$ and $\delta h_n$ in the following calculations.
The collision operator is now spatially dependent and involves the following integrals
\begin{eqnarray}
\oint_R d\mathbf{x}'\delta g_n(\mathbf{x}')&=&2\pi RJ_0(kR)\delta g_n(\mathbf{x})\nonumber\\
\label{BESSEL1}
\int_R d\mathbf{x}'\delta g_n(\mathbf{x}')&=&\frac{2\pi R}{k}J_1(kR)\delta g_n(\mathbf{x}),
\end{eqnarray}
which lead to
\begin{eqnarray}
\oint_R d\mathbf{x}'f(\theta,\mathbf{x}')&=&2\pi R\,\bar{f}+2\pi RJ_0(kR)\,\delta f(\theta,\mathbf{x})\nonumber\\
\label{BESSEL2}
\int_R d\mathbf{x}'f(\theta,\mathbf{x}')&=&\pi R^2\bar{f}+\frac{2\pi R}{k}J_1(kR)\,\delta f(\theta,\mathbf{x}).
\end{eqnarray}
where $J_0$ and $J_1$ are the Bessel functions of the first kind.
%%%%%
Note, the integrals $\oint_R d\mathbf{x}'$ and $\int_R d\mathbf{x}'$ integrate over the circumference of and the area inside the circle with radius $R$ centered around $\mathbf{x}$ respectively. Therefore these integrals are still spatially dependent.
%%%%%  
The line integrations in Eqs. (\ref{BESSEL1}, \ref{BESSEL2}) are related to the ring integral of Eq. (\ref{COLLISION2}) as, 
\begin{equation}
\oint_{R} d\mathbf{x}' \equiv
\lim_{\Delta R\rightarrow 0}\int_{R} {d{\bf x}'\over \Delta R}\,.
\end{equation}
Therefore,
%%%%%
after replacing $\sum_{j=0}^\infty\int_{R(j)} d\mathbf{x}_2$ by the integral $\int_0^\infty dR\oint_R d\mathbf{x}'$,
%%%%%
we have
%\begin{eqnarray}
%&&f(\theta,\mathbf{x}+\tau\mathbf{v},t+\tau)\nonumber\\
%&&=\frac{1}{(M-2)!}\int_{-\eta/2}^{\eta/2}{d\xi\over \eta} \int d\tilde{\theta}_1\ldots %d\tilde{\theta}_M
%\delta(\theta-\xi-\Phi_1)f(\tilde{\theta}_1,\mathbf{x},t)\nonumber\\
%&&\times
%\int_0^{\infty}dR \Big\{\exp{\left[-\pi R^2 \rho_0-\frac{2\pi R}{k}J_1(k,R)\delta\rho(\mathbf{x})\right]}\nonumber\\
%&&\times
%\left[2\pi R\bar{f}(\tilde{\theta}_2)+2\pi R J_0(kR)\delta f(\tilde{\theta}_2,\mathbf{x})\right]\nonumber\\
%&&\times
%\left[\pi R^2\bar{f}(\tilde{\theta}_3)+\frac{2\pi R}{k} J_1(kR)\delta f(\tilde{\theta}_3,\mathbf{x})\right]\nonumber\\
%&&\times \ldots\nonumber\\
%&&\times
%\left[\pi R^2\bar{f}(\tilde{\theta}_M)+\frac{2\pi R}{k} J_1(kR)\delta f(\tilde{\theta}_M,\mathbf{x})\right]\Big\},
%\end{eqnarray}

\begin{eqnarray}
&&f(\theta,\mathbf{x}+\tau\mathbf{v},t+\tau)\nonumber\\
&&=\frac{1}{(M-2)!}\int_{-\eta/2}^{\eta/2}{d\xi\over \eta}\nonumber\\
&& \int d\tilde{\theta}_1\ldots d\tilde{\theta}_M
\delta(\theta-\xi-\Phi_1)f(\tilde{\theta}_1,\mathbf{x},t)\nonumber\\
&&\times
\int_0^{\infty}dR \Bigg\{\exp{\left[-\pi R^2 \rho_0-\frac{2\pi R}{k}J_1(k,R)\delta\rho(\mathbf{x})\right]}\nonumber\\
&&\times
\left[2\pi R\bar{f}(\tilde{\theta}_2)+2\pi R J_0(kR)\delta f(\tilde{\theta}_2,\mathbf{x})\right]\nonumber\\
&&\times\prod_{i=3}^M
\left[\pi R^2\bar{f}(\tilde{\theta}_i)+\frac{2\pi R}{k} J_1(kR)\delta f(\tilde{\theta}_i,\mathbf{x})\right]\Bigg\}.
\end{eqnarray}
Expanding the exponential function in linear order
\begin{eqnarray}
\nonumber
&&\exp{\left[-\pi R^2 \rho_0-\frac{2\pi R}{k}J_1(k R)\delta\rho\right]} \\
\label{eq.ApproxRho}
& =&         
e^{-\pi R^2 \rho_0}\left(1-\frac{2\pi R}{k}J_1(k R)\delta\rho\right)+O(\delta \rho^2)
\end{eqnarray}
and integrating over the collision radius $R$, we arrive at
\begin{eqnarray}
&&f(\theta,\mathbf{x}+\tau\mathbf{v},t+\tau)\nonumber\\
\nonumber
&&=\frac{1}{\rho_0^{M-1}}\frac{1}{\eta}\int_{-\eta/2}^{\eta/2}d\xi \\
&&\int d\tilde{\theta}_1\ldots d\tilde{\theta}_M
\delta(\theta-\xi-\Phi_1)
\mathcal{F}(\tilde{\theta}_1,\tilde{\theta}_2,\ldots\tilde{\theta}_M)
\label{eq.Enskog_MF_2}
\end{eqnarray}
where
\begin{eqnarray}
\nonumber
&&\mathcal{F}(\tilde{\theta}_1,\tilde{\theta}_2,\ldots\tilde{\theta}_M) \\
&&= 
\bar{f}_1 \; \bar{f}_2 \; \bar{f}_3 \ldots \bar{f}_M \left[1- (M-1)\frac{\delta g_0}{g_0}\;{}_1\mathrm{F}_1(M,2,z)\right]\nonumber\\
&&+ \delta{f}_1 \; \bar{f}_2 \; \bar{f}_3 \ldots \bar{f}_M\nonumber\\
&&+ \bar{f}_1 \; \delta f_2 \; \bar{f}_3 \ldots \bar{f}_M \;{}_1\mathrm{F}_1(M-1,1,z)\nonumber\\
&&+ \bar{f}_1 \; \bar{f}_2 \; \delta f_3 \ldots \bar{f}_M \;{}_1\mathrm{F}_1(M-1,2,z)\nonumber\\
&&+ \ldots \nonumber\\
&&+ \bar{f}_1 \; \bar{f}_2 \; \bar{f}_3 \ldots \delta f_M \;{}_1\mathrm{F}_1(M-1,2,z)\nonumber\\
&&+ \mathcal{O}(\delta^2).
\label{eq.Enskog_MF_2_kernel}
\end{eqnarray}
The abbreviations $\bar{f}_j$ and $\delta f_j$ stand for $\bar{f}(\tilde{\theta}_j)$ and $\delta f(\tilde{\theta}_j,\mathbf{x})$ respectively, and ${}_1\mathrm{F}_1(a,b,z)$ is the confluent hypergeometric function with argument $z=-k^2/(4\pi \rho_0)$, see Appendix A. The nonlinear perturbations (denoted as $\mathcal{O}(\delta^2)$) will be neglected in the following. % calculations.
Since the collision integral (\ref{eq.Enskog_MF_2}) is symmetric under permutation of the pre-collision angles $\tilde{\theta}_i$, the integrand can be written as
%
%\begin{eqnarray}
%&&\mathcal{F}(\tilde{\theta}_1,\tilde{\theta}_2,\ldots\tilde{\theta}_M)
%= \\ 
%&&\bar{f}_1 \; \bar{f}_2 \; \bar{f}_3 \ldots \bar{f}_M \left[1- (M-1)\frac{\delta g_0}{g_0}\;{}_1\mathrm{F}_1(M,2,z)\right]\nonumber\\
%&+& \delta{f}_1 \; \bar{f}_2 \; \bar{f}_3 \ldots \bar{f}_M\nonumber\\
%\nonumber
%&\times&
%\left[1+{}_1\mathrm{F}_1(M-1,1,z)+(M-2)\;{}_1\mathrm{F}_1(M-1,2,z)\right]\,.
%\end{eqnarray}
%
\begin{eqnarray}
&&\mathcal{F}(\tilde{\theta}_1,\tilde{\theta}_2,\ldots\tilde{\theta}_M)
\nonumber\\ 
&&=\left(\prod_{i=1}^M\bar{f}_i \right) \left[1- (M-1)\frac{\delta g_0}{g_0}\;{}_1\mathrm{F}_1(M,2,z)\right]\nonumber\\
&&+ \delta{f}_1  \left(\prod_{i=2}^M\bar{f}_i \right) 
\Big[1+{}_1\mathrm{F}_1(M-1,1,z) \nonumber\\
&&+(M-2)\;{}_1\mathrm{F}_1(M-1,2,z)\Big].
\label{eq.Enskog_MF_2_kernel_2}
\end{eqnarray}
%%%%%
Note, the integrand Eq. (\ref{eq.Enskog_MF_2_kernel_2}) is the general form for $M$-particles collision.
For the special case, $M=2$, where there is no particle inside the inner circle, this  simplifies to
%$
%\mathcal{F}(\tilde{\theta}_1,\tilde{\theta}_2)= 
%\bar{f}_1 \; \bar{f}_2 \left[1- \frac{\delta g_0}{g_0}\;{}_1\mathrm{F}_1(2,2,z)\right]
%+\delta{f}_1 \; \bar{f}_2  
%+ \bar{f}_1 \; \delta f_2  \;{}_1\mathrm{F}_1(1,1,z)
%$.
$
\mathcal{F}(\tilde{\theta}_1,\tilde{\theta}_2)= 
\bar{f}_1 \; \bar{f}_2 \left(1- \frac{\delta g_0}{g_0}\;e^z\right)
+\delta{f}_1 \; \bar{f}_2 (1+e^z) 
$,
where ${}_1\mathrm{F}_1(a,a,z)=e^z$ has been applied.
%%%%%
\subsection{Fourier expansion of the collision integral}

The strategy to derive  
the growth rate $\omega(\vec{k})$ is to express both sides of Eq. (\ref{eq.Enskog_MF_2}) in terms of the 
angular Fourier coefficients $C_n({\bf x},t)$ and $H_n({\bf x},t)$, in analogy to the homogeneous case, Eq. (\ref{FOURIERCOLL}).
By equating the different expressions for $C_n$ and $H_n$ obtained from the left hand and the right hand side of Eq.(\ref{eq.Enskog_MF_2}),
a matrix equation for the perturbations $\delta g_k$ and $\delta h_k$ is constructed.
The dispersion relation $\omega({\bf k})$
follows from demanding that there is a nontrivial solution. 

To check the validity of our approach, in particular expansion Eq. (\ref{eq.ApproxRho}), 
we first calculate the zero mode $C_0$ of the collision integral on the r.h.s. of Eq. (\ref{eq.Enskog_MF_2}), namely

\begin{eqnarray}
C_0&=&\frac{1}{2\pi}\int_0^{2\pi}I[f(\tilde{\theta},\mathbf{x},t)]\;d\theta \\
\nonumber
&=& g_0 
+\delta g_0\left[1+{}_1\mathrm{F}_1(M-1,1,z) \right. \\
\nonumber
&+&\left. (M-2)\;{}_1\mathrm{F}_1(M-1,2,z) 
-(M-1)\;{}_1\mathrm{F}_1(M,2,z)\right]
\end{eqnarray}
The confluent hypergeometric functions cancel according to the identities
\begin{eqnarray}
{}_1\mathrm{F}_1(a,b-1,z)-{}_1\mathrm{F}_1(a+1,b,z)&=&\frac{(a-b+1)z}{b(b-1)} \\
\nonumber
&\times&{}_1\mathrm{F}_1(a+1,b+1,z)\\
\nonumber
{}_1\mathrm{F}_1(a+1,b,z)-{}_1\mathrm{F}_1(a,b,z)&=&\frac{z}{b}\;{}_1\mathrm{F}_1(a+1,b+1,z),
\end{eqnarray}
and one finds
\begin{equation}
\label{RHOINV1}
C_0({\bf x},t)= g_0+\delta g_0({\bf x},t)={1\over 2 \pi} \rho({\bf x},t)
\end{equation}
This is the expected result because the collisions do change the local velocity and other moments but
they do not modify the local density $\rho$.
The local density is the zeroth moment of $f$. 
Thus, the invariance of density under collisions requires
\begin{equation}
\label{RHOINV2}
\rho=\int f(\theta)d\theta= 
\int I[f(\theta)]d\theta=2\pi C_0
\end{equation}
which is exactly what we found in Eq. (\ref{RHOINV1}). 

The non-zero modes of the collision integral are
\begin{eqnarray}
C_{n\neq 0}&=&\tilde{A}_n\int d\tilde{\theta}_1\ldots d\tilde{\theta}_M
\mathcal{F}(\tilde{\theta}_1,\tilde{\theta}_2,\ldots\tilde{\theta}_M)\cos{( n\Phi)},
\nonumber\\
H_{n\neq 0}&=&\tilde{A}_n\int d\tilde{\theta}_1\ldots d\tilde{\theta}_M
\mathcal{F}(\tilde{\theta}_1,\tilde{\theta}_2,\ldots\tilde{\theta}_M)\sin{( n\Phi)}.\nonumber\\
\label{RHS_CH}
\end{eqnarray}
where
\begin{equation}
\tilde{A}_n=\frac{1}{\rho_0^{M-1}}\frac{2}{\pi n \eta}\sin\left(\frac{n \eta}{2}\right).
\end{equation}
For a homogeneous ordered state close to the flocking threshold, one has $g_n\sim\epsilon^n
\sim\left({\eta_C-\eta\over \eta_C}\right)^{n/2}$, where $\epsilon$ measures the relative
distance to the critical point, see Eq. (\ref{DEF_EPS}).
Suppose we want to expand $C_n$ and $H_n$ up to order of $\epsilon^z$, we will need all the integrals
\begin{equation}
\int d\tilde{\theta}_1 \ldots d\tilde{\theta}_M T_{0}(n\Phi)T_1(k_1\tilde{\theta}_1) \ldots T_M(k_M\tilde{\theta}_M) 
\end{equation}
which satisfy $k_1+k_2+\ldots+k_M\leq z$, where $k_i$ is a non-negative integer and  function $T_i(x)$ is either $\sin(x)$ or $\cos(x)$. However, not all the integrals have non-zero value. We found that integrals which do not satisfy the condition, 
$\pm n\pm k_1\pm k_2 \pm\ldots \pm k_M=0$, vanish. Furthermore, if the total number of $\sin$ function inside the integral, including $\sin(n \Phi)$, is odd, the integral also vanishes. 
For the binary collision case, $M=2$, by defining 
\begin{eqnarray}
\begin{array}{c}
K^{ccc}_{npq}\equiv  
\langle\cos{(n\Phi)}
\cos{(p\tilde{\theta}_1)}\cos{(q\tilde{\theta}_2)} \rangle \\[0.5em]
K^{css}_{npq}\equiv 
\langle\cos{(n\Phi)}
\sin{(p\tilde{\theta}_1)}\sin{(q\tilde{\theta}_2)}\rangle \\[0.5em]
K^{scs}_{npq}\equiv 
\langle\sin{(n\Phi)}
\cos{(p\tilde{\theta}_1)}\sin{(q\tilde{\theta}_2)}
\rangle
\end{array}
\end{eqnarray}
with $\langle\ldots\rangle\equiv\int_0^{2\pi} d\tilde{\theta}_1\int_0^{2\pi}d\tilde{\theta}_2/(2\pi)^2$,
we have for odd $n$,
\begin{eqnarray}
\nonumber
K^{ccc}_{npq}&=&(+a_1+a_2+a_3)/(2\pi) \\
\nonumber
K^{css}_{npq}&=&(-a_1+a_2+a_3)/(2\pi) \\
\nonumber
K^{scs}_{npq}&=&(+a_1-a_2+a_3)/(2\pi)  \\
\nonumber
a_1&\equiv&{i^{p-q-1} \over p-q}\delta_{p+q,n} \\
\nonumber
a_2&\equiv&\frac{i^{p+q-1}}{p+q}\delta_{p-q,n} \\
a_3&\equiv&\frac{i^{p+q-1}}{p+q}\delta_{q-p,n}\,,
\end{eqnarray}
where $i$ is the imaginary unit.
When $n$ is even we find
\begin{eqnarray}
\nonumber
K^{ccc}_{npq}&=&+\frac{1}{4}\delta_{2p,n} \delta_{2q,n} \\
\nonumber
K^{css}_{npq}&=&-\frac{1}{4} \delta_{2p,n} \delta_{2q,n} \\
K^{scs}_{npq}&=&+\frac{1}{4} \delta_{2p,n} \delta_{2q,n}  
\end{eqnarray}

\subsection{Fourier expansion on the left-hand side}
So far we have considered the Fourier expansion of the collision integral, which is the right-hand side of the
 Enskog-like equation (\ref{eq.Enskog_MF_2}). 
On the left-hand side, writing down the Taylor expansion around $(\mathbf{x},t)$, we have
\begin{eqnarray}
& &f(\theta,\mathbf{x}+\tau\mathbf{v},t+\tau)
=\sum_{n=0}^\infty\frac{\tau^n}{n!}\left(\partial_t+v_{\alpha}\partial_{\alpha}\right)^n f(\theta,\mathbf{x},t)
\nonumber\\
\label{TAYLORLEFT1}
&=&\bar{f}+e^{\tau(\omega+i\mathbf{k}\cdot\mathbf{v})}\sum_{q=0}^\infty\Big[\delta g_q \cos{(q\theta)}+\delta h_q \sin{(q\theta)}\Big]
\end{eqnarray}
The wave vector ${\bf k}$ is split into a longitudinal part and a transversal part with respect to the
average direction of the ordered state ${\bf \hat{n}}$, ${\bf k}=k_{||}{\bf \hat{n}}+k_{\perp}{\bf \hat{t}}$.
Since the coordinate system was chosen with normal direction ${\bf \hat{n}}=(1,0)$ and transversal direction 
${\bf \hat{t}}=(0,1)$, one finds 
\begin{equation}
\mathbf{k}\cdot\mathbf{v}=v_0(k_{||} \cos\theta+k_{\perp}\sin \theta)
\end{equation}
The identities
\begin{eqnarray}
\label{BESSEL_COS}
e^{ix \cos\theta}&=&J_0(x)+2\sum_{n=1}^\infty J_n(x) i^n \cos{(n\theta)} \\ 
\nonumber         
e^{ix \sin\theta}&=&J_0(x)+2i\sum_{n=1}^\infty J_{2n-1}(x) \sin{(n\theta)} \\
\label{BESSEL_SIN}
&+&2\sum_{m=1}^\infty J_{2m}(x) \cos{(m\theta)}\,,
\end{eqnarray}
are used to express the factor $e^{\tau i\mathbf{k}\cdot\mathbf{v}}$
in Eq. (\ref{TAYLORLEFT1}) in terms of Bessel functions.
Because of simplicity and because 
the strongest instabilities of the original Vicsek model occur in the longitudinal direction, where ${\bf k}=k_{||}{\bf \hat{n}}$,
we restrict ourselves to this case. 
It is straightforward to generalize the following analysis to arbitrary directions of the wave vector.
Using Eq. (\ref{BESSEL_COS}) %and abbreviating $f(\theta,\mathbf{x}+\tau\mathbf{v},t+\tau)$ as $f_{t+\tau}$, 
we rewrite Eq. (\ref{TAYLORLEFT1}) as
\begin{eqnarray}
&&f(\theta,\mathbf{x}+\tau\mathbf{v},t+\tau)
=\sum_{j=0}^\infty g_j\cos j\theta
\nonumber\\
&&+~e^{\tau\omega}\left[J_0(\tau v k)+2\sum_{p=1}^\infty J_p(\tau v k) i^p\cos{(p\theta)}\right]
\nonumber\\
&&\times \sum_{q=0}^\infty\Big[\delta g_q \cos{(q\theta)}+\delta h_q \sin{(q\theta)}\Big]
\end{eqnarray}
This can be further converted into the Fourier series 
\begin{eqnarray}
f(\theta,\mathbf{x}+\tau\mathbf{v},t+\tau)&=&C_0+C_1 \cos\theta+C_2 \cos{(2\theta)} +\ldots\nonumber\\
&+& H_1\sin\theta+H_2\sin{(2\theta)}+\ldots\,,
\end{eqnarray}
and allows us to read off the coefficients $C_n$ and $H_n$.
Since, for non-negative integers $p$, $q$, and $n$ 
\begin{eqnarray}
\nonumber
\langle \cos{(p\theta)} \cos{(q\theta)} \cos{(n\theta)}\rangle 
&=&{\delta_{p+q,n}+\delta_{p-q,n}+\delta_{q-p,n}\over 4} \\[0.5em]
\langle\cos{(p\theta)} \sin{(q\theta)} \sin{(n\theta)}\rangle
&=&{\delta_{p+q,n}-\delta_{p-q,n}+\delta_{q-p,n}\over 4}\nonumber\\ 
\end{eqnarray}
with $\langle\ldots\rangle=\int_0^{2\pi}d\theta/(2\pi)$
we have
\begin{eqnarray}
C_0&=&\frac{1}{2\pi}\int_0^{2\pi} f(\theta,\mathbf{x}+\tau\mathbf{v},t+\tau)~d\theta
\nonumber\\
&=&g_0+e^{\tau\omega}\sum_{p=0}^\infty i^p J_p\,\delta g_p
\nonumber\\[1em]
C_{n\neq 0}&=&\frac{1}{\pi}\int_0^{2\pi} f(\theta,\mathbf{x}+\tau\mathbf{v},t+\tau)\cos{(n\theta)} ~d\theta
\nonumber\\
\label{LHS_C}
&=&g_n+e^{\tau\omega}\Big[J_0\,\delta g_n  \\
&+& \sum_{p=1,q=0}^\infty i^p J_p\,\delta g_q
\left(\delta_{p+q,n}+\delta_{p-q,n}+\delta_{-p+q,n}\right)\Big]
\nonumber\\
&=&g_n+e^{\tau\omega}\sum_{q=0}^\infty\left(i^{|n-q|} J_{|n-q|}+i^{n+q} J_{n+q}\right)\,\delta g_q
\nonumber\\[1em]
H_{n\neq 0}&=&\frac{1}{\pi}\int_0^{2\pi} f(\theta,\mathbf{x}+\tau\mathbf{v},t+\tau)\sin{(n\theta)} ~d\theta
\nonumber\\
\nonumber
&=&e^{\tau\omega}\Big[J_0\,\delta h_n \\
\nonumber
&+&\sum_{p=1,q=0}^\infty i^p J_p\,\delta h_q
 \left(\delta_{p+q,n}-\delta_{p-q,n}+\delta_{-p+q,n}\right)\Big]
\nonumber\\
&=&e^{\tau\omega}\sum_{q=0}^\infty\left(i^{|n-q|} J_{|n-q|}-i^{n+q} J_{n+q}\right)\,\delta h_q
.
\label{eq.LHS_expansions}
\end{eqnarray}
In the following, we show examples of $C_n$ and $H_n$ in the expansion up to order $\epsilon^2$, 
\begin{eqnarray}
C_0&=&
g_0+e^{\tau\omega}[J_0\delta g_0+i J_1 \delta g_1-J_2\delta g_2+\ldots]
\nonumber\\
\nonumber
C_1&=&
g_1+e^{\tau\omega}[2i J_1\delta g_0+(J_0-J_2)\delta g_1 \\
\nonumber
&+&i(J_1-J_3)\delta g_2+\ldots]
\nonumber\\
\nonumber
C_2&=&
g_2+e^{\tau\omega}[-2 J_2\delta g_0+i(J_1-J_3)\delta g_1 \\
\nonumber
&+&(J_0+J_4)\delta g_2+\ldots]
\nonumber\\
H_1&=&
e^{\tau\omega}[(J_0+J_2)\delta h_1+i(J_1+J_3)\delta h_2+\ldots]
\nonumber\\
H_2&=&
e^{\tau\omega}[i(J_1+J_3)\delta h_1+(J_0-J_4)\delta h_2+\ldots].
\end{eqnarray}
Equating the expansion of the left-hand side, Eqs. (\ref{LHS_C}, \ref{eq.LHS_expansions}), to the one of the right-hand side, 
Eq. (\ref{RHS_CH}),
a matrix equation of the following general structure is found,
\begin{eqnarray}
\left( \begin{array}{cccccc}
C_{00} & C_{01} & \cdots & A_{01} & A_{02} &\cdots \\
C_{10} & C_{11} & \cdots & A_{11} & A_{12} &\cdots \\
\vdots  & \vdots  & \ddots & \vdots  & \vdots  &\ddots \\
B_{10} & B_{11} & \cdots & H_{11} & H_{12} &\cdots \\
B_{20} & B_{11} & \cdots & H_{21} & H_{22} &\cdots \\
\vdots  & \vdots  & \ddots & \vdots  & \vdots  &\ddots 
\end{array}\right) 
\left(\begin{array}{c}
 \delta g_0\\
 \delta g_1\\
 \vdots\\
 \delta h_1\\
 \delta h_2\\
 \vdots
\end{array}\right)=0
\label{MATRIX}
\end{eqnarray}
According to Eq. (\ref{G1_W_LINK}), for the case ${\bf k}={\bf k_{||}}=k\hat{x}$ considered here, the $\delta g_i$ represent
{\em longitudinal}
perturbations which describe a change of the magnitude but
typically not of the direction of mean flow.
The $\delta h_i$  stand for {\em transversal} perturbations which 
describe odd variations of the distribution function, 
%correspond to odd perturbations,
$\delta f(\theta)=-\delta f(-\theta),$ and modify the flow direction.
Since we only consider linear perturbations,
the $C_n$ of Eq. (\ref{LHS_C}) are found to only depend on $\delta g_j$ but not on $\delta h_j$. Similarly, $H_n$ depends only on $\delta h_j$.
Therefore, 
%for the case $k=k_{||}$ considered here,
the block matrices $A_{\alpha\beta}$ and $B_{\alpha\beta}$ are zero. 
That means, the $\delta g_j$, 
%That means, the $\delta g_i$ 
%which describe a change of the magnitude but
%not of the direction of mean flow 
are decoupled from the $\delta h_j$.
%  which correspond to odd perturbations,
%$\delta f(\theta)=-\delta f(-\theta),$ and modify the flow direction.

The modes $\delta g_n$ and $\delta h_n$ are neglected for $n\ge n_C$ and the block matrices $C_{\alpha\beta}$
and $H_{\alpha\beta}$ are truncated correspondingly.
This truncation is motivated by the observation that, for any nonzero noise, the angular Fourier modes, $g_n$, of the homogeneous ordered state
decay to zero with increasing mode number $n$. It is plausible to assume that the perturbations of these modes,
$\delta g_n$ and $\delta h_n$, show a similar behavior. Of course, this decay will be quite slow if the noise is much smaller than the critical
noise $\eta_C$. This requires a sufficiently large truncation mode number $n_C$.
Its correct choice is discussed in the following chapter and shown in Fig. \ref{OMEGA_M2_DIFF_ORDER}.

Setting the determinants of both matrices equal to zero leads to $2n_C-1$ different branches of  
the dispersion relation $\omega({\bf k})$. The real part of the growth rate, ${\rm Re}(\omega)$, 
of a few of these branches is plotted in Figs. \ref{OMEGA_M2_ETA0.99} -- \ref{OMEGA_M2_DIFF_ORDER}
for different distances to the threshold and for various partner numbers $M$.

\subsection{Results and discussion}

In order to analyze the dispersion relation we distinguish between 
longitudinal and transversal modes.
The longitudinal modes are shown in the left panels of Figs. \ref{OMEGA_M2_ETA0.99} -- \ref{OMEGA_M7_ETA0.99} and \ref{OMEGA_VICSEK_ETA0.99} and
are characterized by changes in the density $\delta \rho \propto \delta g_0$ and in the $x$-component 
of the average flow $\delta w_x\propto \delta g_1$. These changes are always accompanied by corresponding
perturbations of the 
higher order angular coefficients, $\delta g_2, \delta g_3,\ldots \delta g_{n_C-1}$.
The right panels depict the growth rates of the transversal modes.
% which describe changes in the $y$-component
%of the flow, $\delta w_y\propto \delta h_1$ associated with higher order perturbations, $\delta h_2,\delta h_3,\ldots\delta h_{n_C}$.
%associated
%perturbations of the density $\delta \rho \sim \delta g_0$ and of the 
%higher order angular modes, $\delta g_2, \delta g_3,\ldots$ of the distribution function
%and transversal modes plotted in the right panels.
We further distinguish between
hydrodynamic and kinetic modes.
The hydrodynamic modes correspond to exactly conserved quantities and go to zero for $k\rightarrow 0$.
In Figs. \ref{OMEGA_M2_ETA0.99}  -- \ref{OMEGA_M7_ETA0.99} and \ref{OMEGA_VICSEK_ETA0.99} they are plotted as solid lines.
The kinetic modes, defined as those going to a non-zero value at zero wave number, are given as 
dashed lines. The figures show that there are only two hydrodynamic modes. This is expected
because only two quantities are exactly conserved in the collisions: mass and the individual kinetic energy of every particle since the particles' speed never changes.
A closer look reveals that the hydrodynamic mode depicted in the left 
panels of the figures is a sound mode
which is related to longitudinal changes of x-momentum and density. The hydrodynamic mode in the right 
panels of Figs. \ref{OMEGA_M2_ETA0.99}  -- \ref{OMEGA_VICSEK_ETA0.99}
corresponds to changes of the y-component of the momentum. 
At zero $k$ this mode describes the response to
a small rotation of all particle velocity vectors by the same amount. 
Due to rotational invariance this is a Goldstone-mode which meets no resistance and $\omega$ is zero.

We label the top dashed line in the left panels as a ``pseudo''-hydrodynamic mode because
$\omega(k=0)$ is zero only at the critical point, $\eta=\eta_C$ but is negative away from this point.
This mode is related to the fact that, in general, momentum is not conserved in VM-like models but
at the critical point there is no amplification of  momentum perturbations.
Another way to understand this is to look at the hydrodynamic equation for the momentum of the VM, Eq. (5)  in Ref. \cite{ihle_11}, which has the general shape
\begin{equation}
\partial_t\mathbf{w}=(\Gamma-1) \mathbf{w}+O(w^2\mathbf{w})+\ldots
\end{equation}
where $\Gamma$ is the amplification factor defined in Eq. (\ref{OMEGA_KC1DEF}).
At the critical point, $\Gamma(\eta_C)=1$ and $\mathbf{w}$ is small. Thus, at $\eta=\eta_C$, in linear order in the momentum density $\mathbf{w}$, momentum is conserved.

In the left panels of Figs. \ref{OMEGA_M2_ETA0.99}  -- \ref{OMEGA_M2_ETA0.60} we observe 
that the ``pseudo''-hydrodynamic curve 
drops to lower negative values the further away one is from the critical point.
In Fig. \ref{OMEGA_M2_ETA0.60}, at $\eta=0.6\eta_C$, we are so far away from the critical point  
that this mode has now similar
relaxation rates as the other purely kinetic modes.

The main result of this linear stability analysis is that 
there is no longwave instability. In particular, 
we found that  
$Re(\omega)$ of the sound mode, like all other modes, is always negative at small wave numbers. 
This is in contrast to
the regular VM whose modes we show for comparison in Fig. \ref{OMEGA_VICSEK_ETA0.99}.
For the VM, the sound mode  is clearly unstable at wave numbers below $k_C$, confirming the previous 
result from Ref. \cite{ihle_11} which was based on hydrodynamic equations.
In the right panels of Fig. \ref{OMEGA_VICSEK_ETA0.99} we see that all transversal modes are stable.
More details on the regular VM will be reported elsewhere, \cite{chou_12}.

For small truncation level $n_C$
we did see an instability at higher wavenumbers, for $k\lambda\gtrapprox 2.6$.             
However, when $n_C$ is increased, this region is shifted to even
higher $k\gg 1$ whereas $\omega$ remained unchanged at low wavenumbers, 
see Fig. \ref{OMEGA_M2_DIFF_ORDER}.
This strongly suggests that the short wavelength instability is spurious. It is just a result of neglecting 
higher order terms in Eqs. (\ref{GFORALL}) and/or (\ref{MATRIX}) that can be easily remedified.
This result is also consistent with direct simulations of Ref. \cite{chate_10} which showed no sign of instabilities
at any $k$.
For partner number $M=2$ we investigated how far from the threshold one can go and still have 
linear stability of all modes. 
Figs. \ref{OMEGA_M2_ETA0.99}, \ref{OMEGA_M2_ETA0.90} and \ref{OMEGA_M2_ETA0.60} show
that no instability occurs down to $\eta=0.6\eta_C$. 
We did not go to even lower noise because many more terms of the ground state solution $G_n$ and also
more perturbation modes $\delta g_n$ would be needed to achieve reliable results.
Finally, we were interested in how the large partner numbers seen in experiments \cite{ballerini_08,tegeder_95} modify linear stability.
Calculating $\omega({\bf k})$ for all $M$ between two and seven, see Figs. \ref{OMEGA_M3_ETA0.99}, 
\ref{OMEGA_M7_ETA0.99} and \ref{OMEGA_COMPARE_ALL_M}, it is clear that the situation becomes even better:
the larger $M$ is, the more stable the modes, especially the sound mode, become.

To conclude, at $M=2$ -- $7$ there is no signs of linear instability in the metric-free model at or  
below the flocking threshold.
This supports previous claims based on direct simulations 
\cite{chate_10} that the order-disorder transition in this model is continuous and is not made discontinuous by linear instabilities near the threshold.

%-- INSERT physical interpretation ---
To understand why the long wave-length instability of the regular VM
does not appear in the metric-free version, let us compare the momentum amplification factor $\Gamma$ for both models.
According to Eq. (\ref{OMEGA_KC1DEF}), for the metric-free case, $\Gamma$ depends on the partner number $M$ which is a constant.
Thus $\Gamma$ is the same in regions of low and high particle density. 
This is not the case for the metric VM. The amplification factor (which is defined as $\lambda$ in Eq. (3) of Ref. \cite{ihle_11})
depends on the local number of collision partners $M_R({\bf x})$ which is proportional to the local density.
Analyzing this expression shows that, in the metric case, $\Gamma$ is monotonically increasing with density and at high density scales as
$\Gamma\sim \sqrt{\rho}$. As a result the critical noise $\eta_C$ also increases with density.
A possible explanation for the longitudinal instability of the regular VM goes then as follows:
Assume a spatial region of low density, $\rho_L<\rho_0$. The local critical noise, $\eta_C(\rho_L)$ is small, and hence
this region corresponds to a point in either the disordered part of the phase diagram or to a point which is only slightly below
the flocking threshold. This means, on average, particles go in almost all directions and the macroscopic local velocity ${\bf w}/\rho$  
is small or zero. This is consistent with a small amplification rate $\Gamma\approx 1$
and also with Figs. \ref{FIG3}a) and \ref{RYLAN_ORDER}, which show that the average speed decreases monotonically 
if the flocking threshold is approached from inside the ordered phase. 
In a dense region  with $\rho_D> \rho_0$, the local critical noise, $\eta_C(\rho_D)$ would be significantly larger than $\eta$.
This region is then described by a point deep inside the ordered part of the phase diagram.
Thus, particles would be strongly aligned and the average local speed would be large, consistent with a large $\Gamma$. 
These are exactly the conditions to form a density wave: particles in high density regions are more aligned and ``invade''
regions of lower density where particles perform a slower average motion and thus are ``not organized enough'' to escape from the 
dense crowd coming in.
In the metric-free model, the local density does not couple to the average particle motion in this way. The mechanism for density wave
formation discussed above is absent.
A similar discussion of the absence of instabilities in metric-free models is given in Ref. \cite{peshkov_12}.
In appendix D we relate our approach to the work of Ref. \cite{peshkov_12} and investigate the role of collisional
momentum transfer by considering various limits of the general Enskog-like kinetic equation, Eq. (\ref{COLLISION4}).

%-- END INSERT

\begin{figure}
\begin{center}
\vspace{0cm}
\includegraphics[width=3.1in,angle=0]{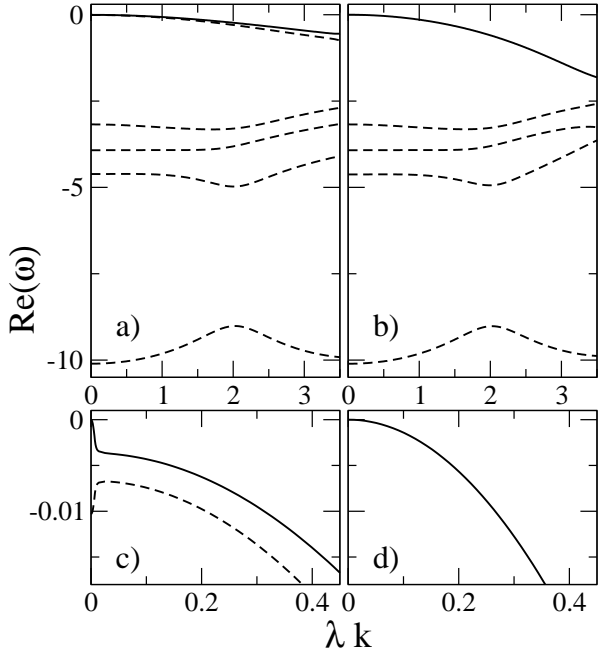}
\caption{Real part of the growth rate $\omega$ as a function of wave number for $M=2$, close to the flocking threshold, at 
$\eta=0.99\eta_C$, calculated
from Eq. (\ref{MATRIX}).
Other parameters: $\Lambda=2$, $\rho_0=0.1$.%$\tau=v_0=1$. 
The ground state solution is accurate up to order $\epsilon^5$.
Perturbations $\delta g_n$, $\delta h_n$ with $n\ge 6$ were neglected. 
Part a) shows solution branches for the determinant equation, $\text{det}({\bf C})=0$, where the block matrix $C_{\alpha\beta}$ is defined
in Eq. (\ref{MATRIX}).
%determinant of the block matrix $C_{\alpha\beta}$, being zero, $\text{det}({\bf C})=0$, see Eq. (\ref{MATRIX}).
These curves describe different {\em longitudinal} excitations where typically all angular perturbation coefficients
$\delta g_0, \delta g_1,\ldots \delta g_n$ are nonzero. 
Within a particular excitation mode and for a given wave number, the coefficients $\delta g_n$ occur in fixed specific ratios to each other.
%The perturbations
%occur in fixed ratios to each other which are specific for a particular excitation.
Part b) shows the real part of the growth rate for different {\em transversal} modes which are composed of the 
$\delta h_1, \delta h_2,\ldots \delta h_n$. 
%Part a) shows the growteigenmodes for the $\delta g_n$ solutions, whereas b) correspond to $\delta h_n$ modes.
Hydrodynamic modes are plotted as solid lines, the kinetic modes are dashed.
Parts c) and d) are just zoomed in versions of a) and b), respectively, to better show the small $k$ behavior.
}
\label{OMEGA_M2_ETA0.99}
\end{center}
\vspace*{-0.5ex}
\end{figure}

\begin{figure}
\begin{center}
\vspace{0cm}
\includegraphics[width=3.1in,angle=0]{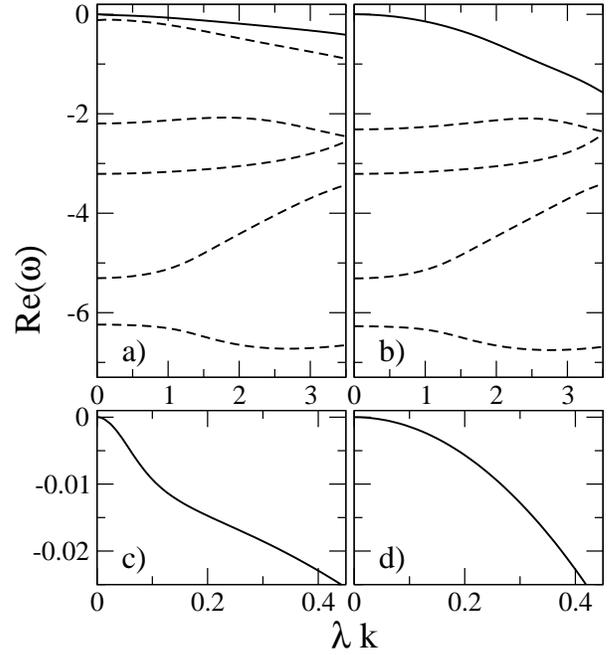}
\caption{Real part of the growth rate $\omega$ as a function of wave number for $M=2$ and $\eta=0.90\eta_C$.
Notation and other parameters are the same as in Fig. \ref{OMEGA_M2_ETA0.99}.
}
\label{OMEGA_M2_ETA0.90}
\end{center}
\vspace*{-0.5ex}
\end{figure}

\begin{figure}
\begin{center}
\vspace{0cm}
\includegraphics[width=3.1in,angle=0]{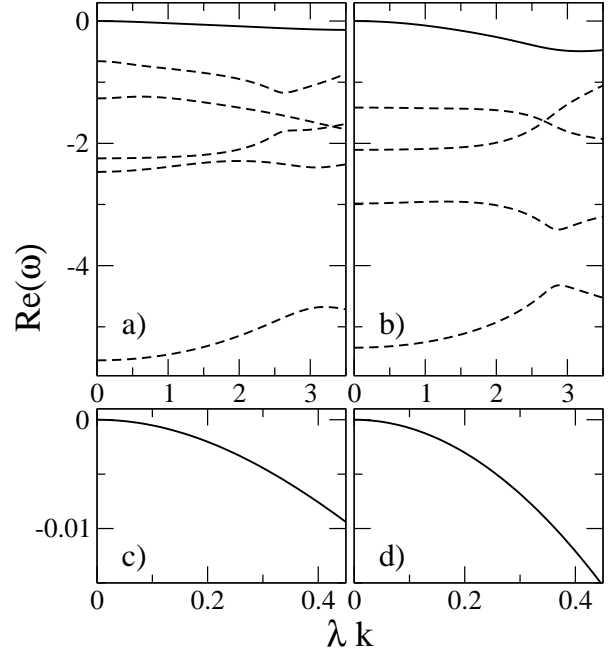}
\caption{Real part of the growth rate $\omega$ as a function of wave number for $M=2$ far from the flocking threshold at $\eta=0.60\eta_C$.
Notation and other parameters are the same as in Fig. \ref{OMEGA_M2_ETA0.99}.
}
\label{OMEGA_M2_ETA0.60}
\end{center}
\vspace*{-0.5ex}
\end{figure}

\begin{figure}
\begin{center}
\vspace{0cm}
\includegraphics[width=3.1in,angle=0]{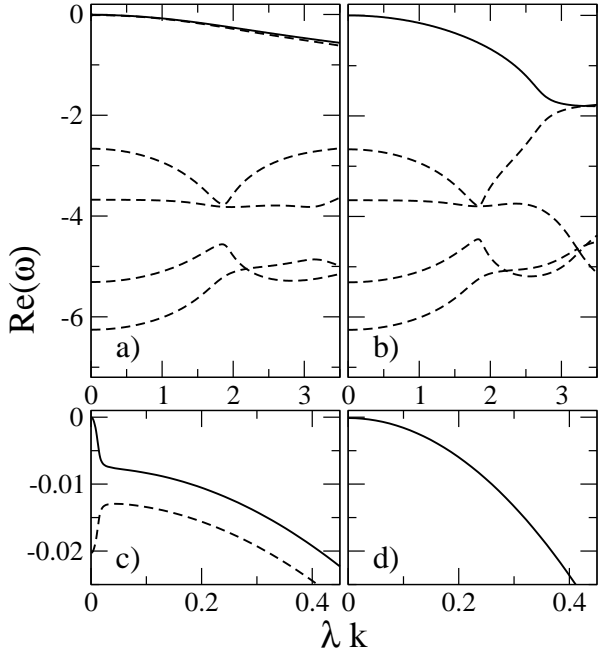}
\caption{Real part of the growth rate $\omega$ as a function of wave number for $M=3$ close to the flocking threshold, at $\eta=0.99\eta_C$.
Other parameters are the same as in Fig. \ref{OMEGA_M2_ETA0.99}.
}
\label{OMEGA_M3_ETA0.99}
\end{center}
\vspace*{-0.5ex}
\end{figure}

\begin{figure}
\begin{center}
\vspace{0cm}
\includegraphics[width=3.1in,angle=0]{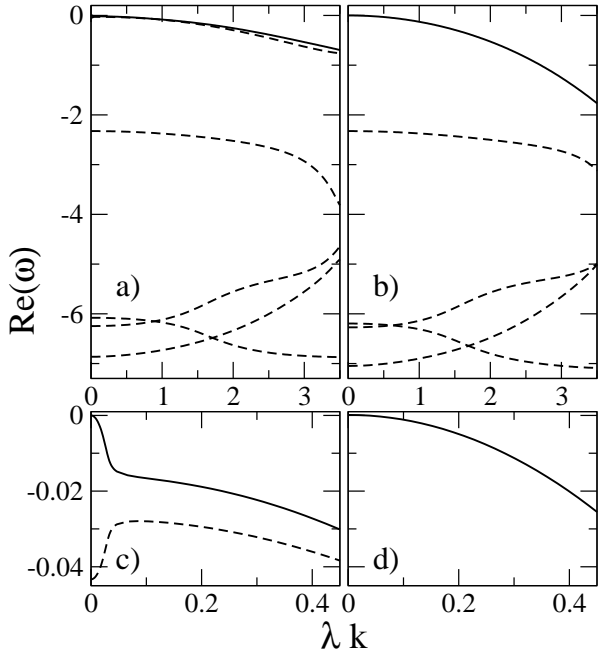}
\caption{Real part of the growth rate $\omega$ as a function of wave number for $M=7$ at $\eta=0.99\eta_C$.
Other parameters are the same as in Fig. \ref{OMEGA_M2_ETA0.99}.
}
\label{OMEGA_M7_ETA0.99}
\end{center}
\vspace*{0.5ex}
\end{figure}

\begin{figure}
\begin{center}
\vspace{0cm}
\includegraphics[width=3.1in,angle=0]{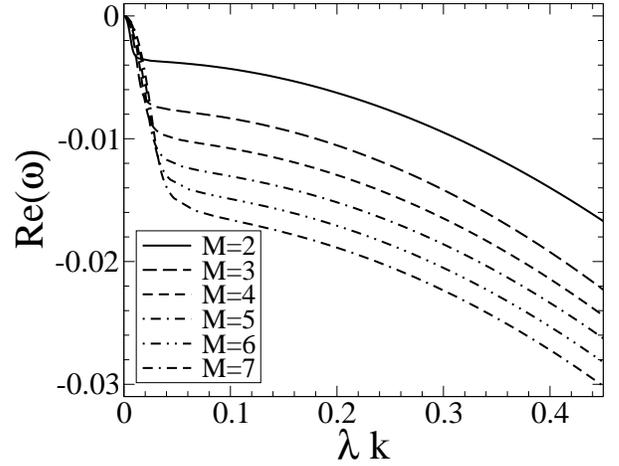}
\caption{Real part of the growth rate $\omega$ of the sound mode for different $M$ at $\eta=0.99\eta_C$.
Other parameters are the same as in Fig. \ref{OMEGA_M2_ETA0.99}.
}
\label{OMEGA_COMPARE_ALL_M}
\end{center}
\vspace*{5ex}
\end{figure}

\begin{figure}
\begin{center}
\vspace*{0.5ex}
\includegraphics[width=3.1in,angle=0]{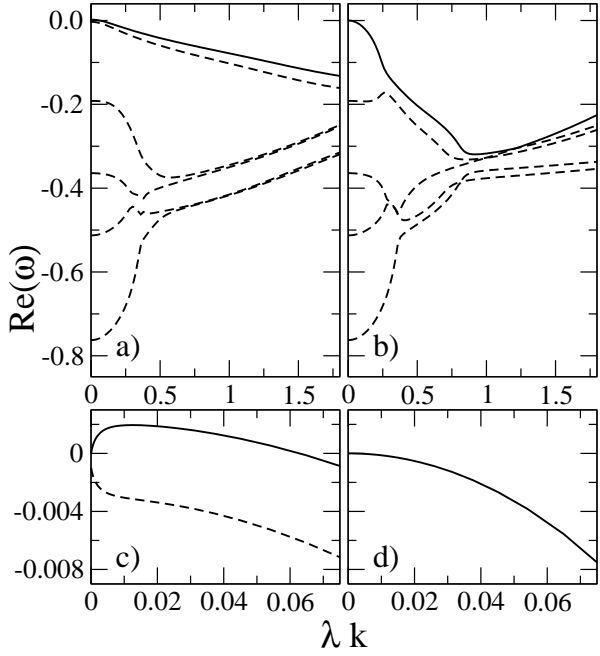}
\caption{Regular Vicsek model: Real part of the growth rate $\omega$ as a function of wave number for low density, $\langle M\rangle=0.1$, \
close to the flocking threshold, at $\eta=0.99\eta_C$.
Note the long wavelength instability in part c).
Other parameters: $\Lambda=\lambda/R=4$. Details will be given elsewhere \cite{chou_12}.
}
\label{OMEGA_VICSEK_ETA0.99}
\end{center}
\vspace*{1.5ex}
\end{figure}

\begin{figure}
\begin{center}
\vspace{0cm}
\includegraphics[width=3.1in,angle=0]{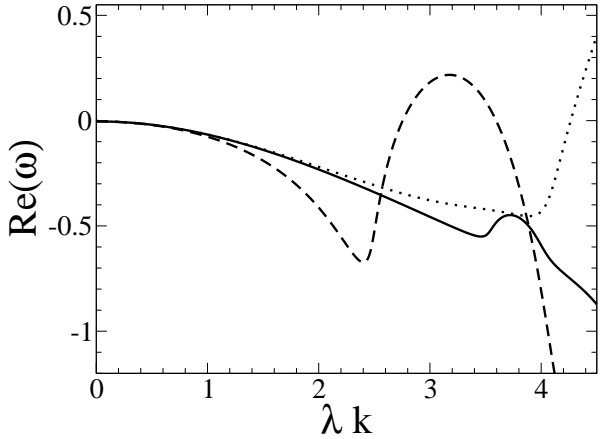}
\caption{Effects of truncating the matrix equation, Eq. (\ref{MATRIX}):
Real part of the growth rate $\omega$ of the {\em slowest} $\delta g_n$ mode as a function of wave number 
for truncation levels $n_C=3$ (dashed line), $n_C=4$ (dotted) and $n_C=5$ (solid line). 
Parameters: $M=2$, $\eta=0.99\eta_C$, $\Lambda=2$, $\rho_0=0.1$.
}
\label{OMEGA_M2_DIFF_ORDER}
\end{center}
\vspace*{-2ex}
\end{figure}

\section{Direct numerical simulation}

\begin{figure}
\begin{center}
\vspace{0cm}
\includegraphics[width=3.2in,angle=0]{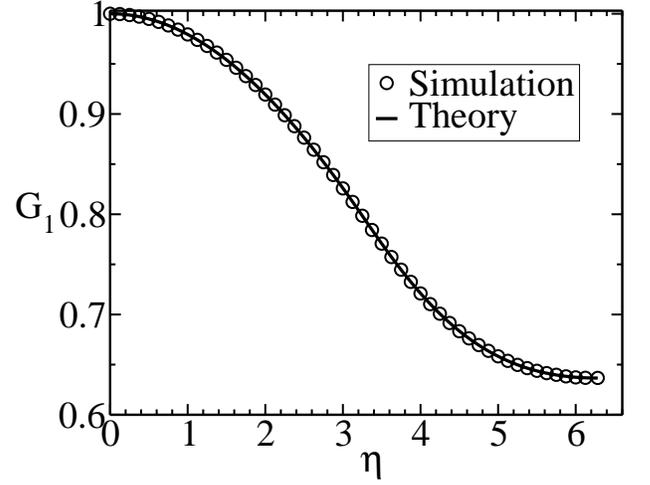}
\caption{
The order parameter $G_1$ as a function of noise for the special case $N=M=2$.
The exact solution, Eq. (\ref{EXACT_NM2}), (solid line) perfectly agrees with the simulation (circles).
}
\label{RYLANN2}
\end{center}
\vspace*{-1ex}
\end{figure}

\begin{figure}
\begin{center}
\vspace{0cm}
\includegraphics[width=3.2in,angle=0]{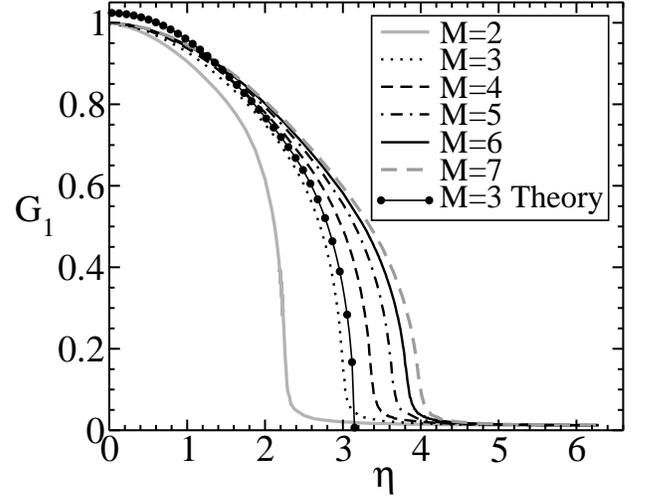}
\caption{
Direct simulations: The order parameter $G_1$ as a function of noise for 
various $M=2\ldots 7$ and at $N=5000$, $\Lambda=\lambda/R_{\text{eff}}=5.66$.
The analytical solution for $M=3$ (filled circles) was obtained from Eq. (\ref{GFORALL}) with 
terms up to order $\epsilon^5$. 
}
\label{RYLAN_ORDER}
\end{center}
\vspace*{-2ex}
\end{figure}

\begin{figure}
\begin{center}
\vspace{0cm}
\includegraphics[width=3.2in,angle=0]{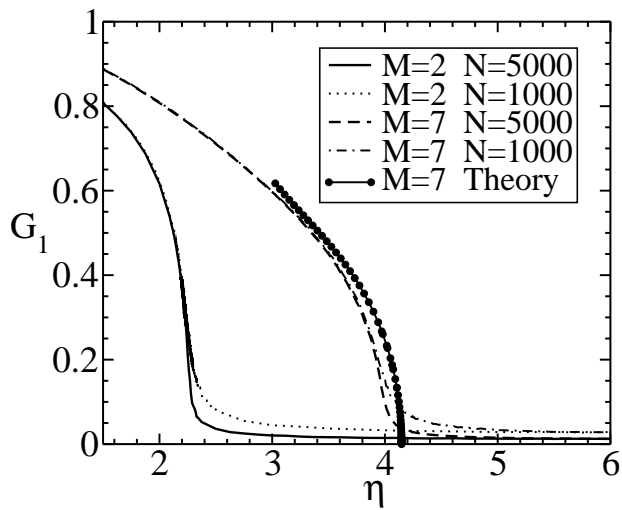}
\caption{
Direct simulations: The order parameter for 
different particle numbers: $N=1000$ and $5000$; $\Lambda=5.66$.
The theoretical result for $M=7$ calculated at order $\epsilon^5$ from Eq.
(\ref{GFORALL}) is shown by the filled circles.
}
\label{RYLAN_DIFFERENTN}
\end{center}
\vspace*{-2ex}
\end{figure}
In order to verify our analytical results
we also performed direct numerical simulations of the model defined in Eq. (\ref{INTERACTION}). 
A quadratic simulation box of size $L\times L$ with periodic boundary conditions is used and seeded
with $N$ particles. Their initial positions and flying directions are chosen at random.
For every particle $i$, the distances to all other particles are measured and 
the $M-1$ particles with the smallest distances are defined as the neighbors of particle $i$.
This differs from the neighboorhood definition of Ref. \cite{chate_10}
by means of a Voronoi-construction.

After the system had relaxed into a stationary state, we performed a time average of the order parameter 
\begin{equation}
\Omega={1\over N v_0}\left|\sum_{i=1}^N{\bf v}_i\right|
\end{equation}
and plotted it as a function of various parameters, see Figs. \ref{RYLAN_ORDER} and 
\ref{RYLAN_DIFFERENTN}.
While typical particle numbers are $N=1000$ and 5000, we first performed a test at $N=M=2$ because an exact
result without the molecular chaos approximation can be derived for the order parameter
\begin{equation}
\Omega= 
  \begin{cases}
   \left( {4\over \eta}\sin{\eta\over 4}\right)^2           & \text{if } \eta \leq \pi \\
   8\left[1+\eta-\pi+\cos{\eta\over 2}\right]/\eta^2  & \text{if } \pi\leq\eta\leq 2\pi 
  \end{cases}
\label{EXACT_NM2}
\end{equation}
see Eqs. (\ref{SMALLETA}, \ref{LARGEETA}) in Appendix B.
As seen in Fig. \ref{RYLANN2} the simulations for $N=2$ are in excellent agreement with this formula.

Another analytical result was obtained for maximum noise strength $\eta=2\pi$ where a mapping to a random walk can be utilized
and for large $N$
\begin{equation}
\Omega(\eta=2\pi)\approx {7\over 8}{1\over \sqrt{N}} 
\end{equation}
is obtained, see Appendix B.
Our simulations were also in excellent agreement with this expression.
The derivation of the phase diagram relied on the approximation of molecular chaos.
For nonzero noise and finite $M$, this approximation
becomes exact at infinite mean free path $\lambda$.
Therefore, the order parameter and the phase diagram were measured at a large ratio $\Lambda=\lambda/R_{\text{eff}}=5.6$ and 
compared with
analytical results in Figs. \ref{RYLAN_ORDER} and \ref{RYLAN_DIFFERENTN}.
In order to investigate the importance of the mean free path (mfp), the critical noise at fixed $M$
but different $\Lambda$ was determined, see Fig. \ref{FIG_DIFF_LAM}.
We found that for both $M=2$ and $M=7$ the influence of the mfp is only negligible if
$\Lambda$ is above one. For smaller mfp's, the differences are significant and are subject to current 
studies.
For example, $\eta_C$ drops by almost a factor of three if $\Lambda$ is reduced from $5.6$ to $0.1$.
\begin{figure}
\begin{center}
\vspace{0cm}
\includegraphics[width=3.2in,angle=0]{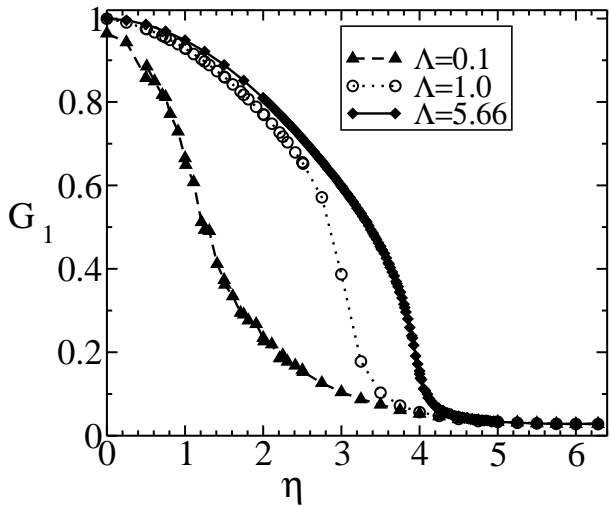}
\caption{
Order parameter for $M=7$ at different mean free paths, $\Lambda=\lambda/R_{\text{eff}}$,
with $N=1000$. The lines just serve as guides to the eye. 
}
\label{FIG_DIFF_LAM}
\end{center}
\vspace*{-0.5ex}
\end{figure}

\begin{figure}
\begin{center}
\vspace{0cm}
\includegraphics[width=3.2in,angle=0]{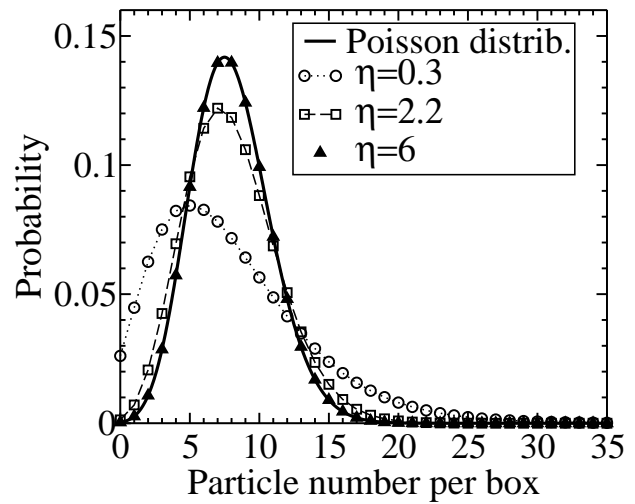}
\caption{
Histogram for different noises at  $M=2$ in comparison with the Poisson distribution (solid line)
as a function of particle number per box, $n$. 
Simulation parameters: $N=5000$, $\Lambda=5.66$, $\langle n \rangle=8$, time average over $25000$ measurements. The dashed and dotted lines are only guides to the eye.
}
\label{RYLANHISTO}
\end{center}
\vspace*{-0.5ex}
\end{figure}

The main assumption of our mean-field theory, the molecular chaos approximation, predicts that the particle number in a given box is
Poisson-distributed \cite{ihle_11}. Hence, the probability to find $n$ particles in a box of area $V$ is given by
\begin{equation}
\label{POISSON}
p_n={\rm e}^{-\langle n \rangle} {\langle n \rangle^n\over n!}
\end{equation}
where $\langle n\rangle=V\rho_0$ is the average particle number in that box.
In order to indirectly test the validity of the molecular chaos approximation, we divided our simulation domain into $25\times 25$ quadratic
boxes such that $\langle n \rangle=8$, and recorded how often a box was occupied by a given particle number $n$.
These measurements were averaged over all boxes and over time. The results for $M=2$ are shown in 
Fig. \ref{RYLANHISTO} 
and compared to an analytic continuation of Eq. (\ref{POISSON}) to real numbers where $n!$ was replaced by the gamma function, $\Gamma(n+1)$.
For large noise $\eta=6$, which is far beyond the order-transition threshold $\eta_{C,sim}=2.27$, the histogram shows perfect
agreement with the Poisson distribution. This confirms our expectation that the Molecular Chaos approximation should be valid in the 
disordered phase. However, at noise $\eta=2.2$, that is in the ordered phase but only $3\%$ below the threshold, a clear deviation from
the Poisson distribution occurs and the maximum of the curve lies about $20\%$ below the Poisson curve.
This is interesting because despite this deviation the shape of the order parameter curve and the critical noise value differ much less
from the mean field predictions.
Finally, at very low noise, $\eta=0.3=0.13\eta_{C,sim}$, the particle number distribution is much wider than the Poisson distribution.
In particular, it is much more probable to find empty boxes and boxes occupied 
with more than three times the average number.
We see that even though no density bands are observed as in the regular VM, there is still anomalously large density fluctuations.

\section{Conclusion}
We have presented a detailed, systematic derivation of a kinetic theory
for a model of self-propelled particles with metric-free interactions.
This discrete-time model has 
genuine multi-body interactions and was introduced in Ref. \cite{ballerini_08}.
The sole approximation in the derivation was the assumption of Molecular chaos which we used to reduce an exact Master-equation 
for all particles to an equation for the one-particle density.

This novel Enskog-type kinetic equation, Eq. (\ref{COLLISION4}), is one of the main results of this paper. 
Using this equation, the transition from a disordered state to a homogeneous state of collective motion
was studied for various numbers of interaction partners, $M$. 
We calculated the phase diagram and the order parameter 
analytically as well as numerically and also performed direct simulations. 
We found excellent agreement within a few percent between theory and simulation
as long as the mean free path is at least several times larger than the effective interaction radius. 
In order to test the validity of the molecular chaos approximation, 
we
recorded particle density histograms and measured how lowering the mean free path affects the phase diagram.

Simulations of the regular Vicsek-model showed that the flocking transition becomes discontinuous once
the system size is beyond a certain critical length \cite{chate_04_08}.
This observation has been linked to a long wave length instability of the ordered phase 
right below the flocking threshold \cite{ihle_11,bertin_06_09}.
In this paper, 
we have performed a linear stability analysis of the ordered state of the metric-free model 
in order to investigate the nature of the flocking transition in the presence of topological interactions. 
This was done by directly imposing perturbations into the kinetic equation 
without 
%making a detour via 
first deriving
hydrodynamic equations. 
Such derivations are very tedious for models with a finite time step and multi-body interactions, see Refs. \cite{ihle_11,malev_99,ihle_09}.
If one is only interested in the linear stability of a certain phase it is more convinient to use the kinetic equations directly.
%Obtaining accurate expressions for the transport coefficients ydrodynamic equations is very tedious
An additional advantage is that by not imposing closure of the kinetic equations at a predetermined low level
and refraining from gradient expansions of any kind, higher accuracy and a larger range of validity of the stability analysis can be achieved.
The main result is that for all partner numbers $2\leq M\leq 7$ we tested, all modes are stable right next to the flocking
threshold. For select $M$ we verified that even very far from threshold, no linear instabilities occur.
This result is consistent with direct simulations of the metric-free model where no high density 
bands -- a sign of instability -- were observed 
and where the flocking transition was found to be continuous.
While our results do not come as a big surprise, they do rule out the possibility of a linearly unstable but nonlinearly stable ordered state.
This would
lead to an inhomogeneous ordered state which would be hard to identify in a direct simulation if the inhomogeneity is small.

The existence of a longitudinal instability can be related to the different parameters that span the 
phase diagram:  
In models with metric interactions, the critical noise depends on local density. 
Thus, different spatial regions can be characterized by different phase space distances,
$\Delta \eta=\eta_C(\rho_{\rm local})-\eta$,
to the flocking threshold. In metric-free models all regions are at the same distance to the threshold.
With the additional facts that the order parameter is monotonically increasing with $\Delta \eta$ and that the critical noise
of the metric VM is increasing with density, an intuitive understanding of the occurence of density waves in the metric models 
and of their absence in topological models can be obtained.

It remains an open question how to systematically go beyond the approximation of molecular chaos in order to improve the results at low mfp. 
One possibility is to derive additional noise terms in the hydrodynamic or kinetic equations 
along the lines of 
Refs. \cite{peliti_85,tailleur_08,thompson_11,pierre_98}. Work in this direction is in progress.

We did not derive hydrodynamic equations for the metric-free model because it is fairly obvious
that they must have exactly the same shape as Eq. (5) of Ref. \cite{ihle_11} 
which was derived for the regular Vicsek model.
This is because both models have the same symmetries (rotational and translational), no Galilean invariance, and the same set of conserved quantities (mass and kinetic energy).
The Chapman-Enskog expansion of Ref. \cite{ihle_11}, which is basically a gradient expansion, gave all
possible terms allowed by the symmetries and by the order of the expansion.
Therefore, the metric-free model has no other ``choice'' than picking the same terms just with different coefficients.

We also identified the {\em ad hoc} Boltzmann-like collision integral of Ref. \cite{peshkov_12} as the zero wave number or infinite $\Lambda$ limit
of our theory for the special case of $M=2$, see Appendix D.
This was used to quantify the relevance of ``collisional momentum transfer'' to the linear stability of the ordered phase.
Finally, the kinetic formalism presented in this paper might also be useful to treat 
other ``exotic'', multi-body
interaction rules which are often postulated in ecological modelling of animals \cite{animal_flocks}, 
human crowds \cite{helbing_01}, and interacting robots \cite{jadba_03,turgut_08,shen_04}.

\section{Acknowledgments}
Support
from the National Science Foundation under grant No.
DMR-0706017 
is gratefully acknowledged.
One of us would like to thank C. Huepe, J. Tailleur and F. Peruani
for stimulating discussions.
TI thanks F. J{\"u}licher for his hospitality at the Max-Planck Institute
for Complex Systems in Dresden, where part of this study was performed.

\section*{Appendix A: Hypergeometric functions}
Using the expansion of Bessel function of the first kind 
\begin{equation}
J_n(x)=\sum_{s=0}^\infty\frac{(-1)^s}{s!(n+s)!}\left(\frac{x}{2}\right)^{n+2s}
\end{equation}
and the integral
\begin{equation}
\int_0^\infty x^me^{-a x^2}dx=\frac{\Gamma[(m+1)/2]}{2a^{(m+1)/2}}\text{ , for $a>0$ and $m\geq 0$},
\end{equation}
we have
\begin{eqnarray}
&&\int_0^\infty x^m J_n(x) e^{-a x^2} dx
\nonumber\\
&&=\sum_{s=0}^\infty\frac{(-1)^s}{s!(n+s)!}\frac{1}{2^{n+2s+1}}\Gamma\left(\frac{m+n+2s+1}{2}\right)a^{-\frac{m+n+2s+1}{2}}
\nonumber
\end{eqnarray}
Setting $l\equiv (m+n+1)/2$, the series becomes
\begin{eqnarray}
&&=\frac{1}{2^{n+1}a^l}\sum_{s=0}^\infty
\frac{(l+s-1)(l+s-2)\ldots l(l-1)!}{(n+s)(n+s-1)\ldots(n+1)n!s!}\left(\frac{-1}{4a}\right)^s
\nonumber\\
&&=\frac{1}{2^{n+1}a^l}\frac{\Gamma(l)}{\Gamma(n+1)}
~{}_1F_1\left(l,n+1,-\frac{1}{4a}\right),
\end{eqnarray}
where
\begin{equation}
{}_1F_1(a,b,z)=\sum_{s=0}^\infty\frac{(a)_s}{(b)_s}\frac{z^s}{s!}
\end{equation}
is the confluent hypergeometric function in the notation of Pochhammer symbol $(x)_n=x(x+1)\ldots(x+n-1)$.
If the argument of Bessel function rescales as $x\rightarrow kx$, it is easy to show that the integral becomes
\begin{eqnarray}
&&\int_0^\infty x^m J_n(kx) e^{-a x^2} dx
\nonumber\\
\nonumber
&&=\frac{1}{2^{n+1}}\frac{k^n}{a^{(m+n+1)/2}}\frac{\Gamma\left(\frac{m+n+1}{2}\right)}{\Gamma(n+1)} \\
&&\times~{}_1F_1\left(\frac{m+n+1}{2},n+1,-\frac{k^2}{4a}\right).
\end{eqnarray}
%%%%%

\section*{Appendix B: Exact solutions}

\subsection{Case N=2}

For $M=N$ every particle is neighbor to every other particle, which can be analytically exploited.
We consider the special case $N=M=2$.
The order parameter is expressed as, 
\begin{equation}
\label{ORDER_B}
\Omega={1\over 2}\langle | {\bf \hat{n}}_1+{\bf \hat{n}}_2|\rangle\,,
\end{equation}
where ${\bf \hat{n}}_i$ are the normalized velocity vectors {\em after} a collision.
The angular brackets denote the average over the two uncorrelated angular noises $\xi_1$ and $\xi_2$.
The average angle $\Phi$ is the same for both particles and is therefore irrelevant
for the order parameter. We choose it to be zero.
In this case, ${\bf \hat{n}}_i=(\cos \xi_i, \sin \xi_i)$, and we have
\begin{eqnarray}    
\nonumber
\Omega&=&{1\over \eta^2} \int_{-\eta/2}^{\eta/2}d\xi_1\int_{-\eta/2}^{\eta/2}d\xi_2 
I(\xi_1,\xi_2) \\
\nonumber
I(\xi_1,\xi_2)&=&{1\over 2}\sqrt{(\cos\xi_1+\cos\xi_2)^2+(\sin\xi_1+\sin\xi_2)^2} \\
\nonumber
              &=&{1\over 2}\sqrt{2 (1+\cos(\xi_1-\xi_2)} \\
\nonumber
              &=&\left| \cos{\xi_1-\xi_2 \over 2}\right| \\
\label{N2_FORMULA1}
              &=&\left| \cos{\xi_1\over 2} \cos{\xi_2\over 2}+\sin{\xi_1\over 2}\sin{\xi_2\over 2}\right|
\end{eqnarray}
For $\eta\le\pi$ the integrations are straightforward since $|\xi_1-\xi_2|\le \pi$ and give
\begin{equation}
\label{SMALLETA}       
\Omega=\left( {4\over \eta}\sin{\eta\over 4}\right)^2
\end{equation}
For $\pi\le\eta\le 2\pi$ new variables are introduced,
\begin{eqnarray}
\alpha_1&=&\xi_1-\xi_2 \\
\alpha_2&=&\xi_2\,,
\end{eqnarray}
and the integration area has to be split into several 
domains
in order to correctly treat the absolute value in Eq. (\ref{N2_FORMULA1}).
This gives
\begin{eqnarray}
\nonumber
\Omega&=&{1\over \eta^2}\Big[ 
 \int_{-\eta}^{-\pi}(-A_1)\,d\alpha_1+
 \int_{-\pi}^{-\eta/2}A_1\,d\alpha_1 \\
\nonumber
&+&\int_{\eta/2}^{\pi}(A_2)\,d\alpha_1+
 \int_{\pi}^{\eta}(-A_2)\,d\alpha_1 \\
&+&
\int_{-\eta/2}^0 A_3\,d\alpha_1
+\int_{0}^{\eta/2} A_4\,d\alpha_1
\Big]
\end{eqnarray}
with
\begin{eqnarray}
\nonumber
A_1&=&\int_{-\eta/2-\alpha_1}^{\eta/2}(-c)\,d\alpha_2 \\
\nonumber
A_2&=&\int_{-\eta/2}^{-\alpha_1+\eta/2}(-c)\,d\alpha_2 \\
\nonumber
A_3&=&\int_{-\eta/2-\alpha_1}^{\eta/2} c\,d\alpha_2 \\
\nonumber
A_4&=&\int_{-\eta/2}^{\eta/2-\alpha_1} c\,d\alpha_2
\end{eqnarray}
where $c\equiv \cos(\alpha_1/2)$.

Integrating over $\alpha_2$ yields,
\begin{eqnarray}
\nonumber
\Omega&=&{1\over \eta^2}\Big[ 
\int_{-\eta}^{-\pi}(-c)(\eta+\alpha_1)\,d\alpha_1+
\int_{-\pi}^{-\eta/2}(+c)(\eta+\alpha_1)\,d\alpha_1 \\
\nonumber
&+&
\int_{\eta/2}^{\pi}(+c)(\eta-\alpha_1)\,d\alpha_1+
\int_{\pi}^{\eta}(-c)(\eta-\alpha_1)\,d\alpha_1 \\
&+&
2\int_{0}^{\eta/2}(+c)(\eta-\alpha_1)\,d\alpha_1
\Big]\,.
\end{eqnarray}
After integration we obtain for $\pi\le \eta\le 2\pi$:
\begin{equation}
\label{LARGEETA}
\Omega={8\over \eta^2}\left[1+\eta-\pi+\cos{\eta\over 2}\right]
\end{equation}
At $\eta=\pi$ both expressions, Eq. (\ref{SMALLETA}) and (\ref{LARGEETA})
match at $\Omega=8/\pi^2\approx 0.81$.
This point, $\eta=\pi$, is also the turning point of the order parameter curve, Fig. \ref{RYLANN2}.

At the largest noise value, $\eta=2\pi$, we find
$\Omega=2/\pi\approx 0.6366$. 
It is interesting to note that the approximation for large $N$ and $\eta=2\pi$, Eq. (\ref{LARGE_NOISE_N}), even works well
for this $N=2$ case, since $7/(8 \sqrt{2})=0.6187$ is only 3\% smaller than
the exact result.

\subsection{Case ${\bf N \gg 1}$, ${\bf \eta=2\pi}$}

The order parameter formula, Eq. (\ref{ORDER_B}) is generalized to $N$ particles,
\begin{eqnarray}
\nonumber
\Omega&=&{1\over N}\Big\langle \Big| \sum_{i=1}^N{\bf \hat{n}}_i\Big|\Big\rangle\\
\nonumber
&=&\Bigg\langle\sqrt{\Big[\sum_i \cos \xi_i\Big]^2
+\Big[\sum_i \sin \xi_i\Big]^2}\,\Bigg\rangle \\
\label{ORDER_LARGE_N}
\Big\langle\ldots\Big\rangle&\equiv & \prod_{i=1}^N
\left(
{1\over \eta}\int_{-\eta/2}^{\eta/2}d\xi_i
\right)
\end{eqnarray}
Using $\cos^2\xi_i+\sin^2\xi_i=1$ 
the terms inside the square root can be reordered 
with the result,
\begin{eqnarray}
\nonumber
\Omega&=&{1\over \sqrt{N}}
\Bigg\langle
\sqrt{1+{A+B\over N}}
\Bigg\rangle \\
\nonumber
A&\equiv& 2\sum_i \sum_{j>i} c_i c_j \\
\label{APPB_SQRT1}
B&\equiv& 2\sum_i \sum_{j>i} s_i s_j 
\end{eqnarray}
Since $c_i\equiv\cos\xi_i$ and $s_i\equiv\sin\xi_i$ vary between $-1$ and $1$ the terms $A$ and $B$
will be smaller than $N$ for most realizations when the particle number $N$ is large.
We therefore attempt a Taylor expansion of the square root in Eq. (\ref{APPB_SQRT1})
whose validity can be checked {\em a posteriori}.
We obtain,
\begin{eqnarray}
\nonumber
\Omega&=&{1\over \sqrt{N}}\Big(1+{\langle A\rangle+\langle B\rangle\over 2 N} \\
\label{APPB_SQ}
&-&{1\over 8 N^2}[\langle A^2\rangle+\langle B\rangle^2+2\langle A B \rangle]
+\ldots
\Big)
\end{eqnarray}
Since the angular noises $\xi_i$ are uncorrelated one finds
\begin{eqnarray}
\nonumber
\langle A\rangle=\langle B\rangle=\langle AB\rangle&=&0 \\
\langle A^2\rangle=\langle B^2\rangle&=& {N(N-1)\over 2}
\end{eqnarray}
Substituting into Eq. (\ref{APPB_SQ}) gives
\begin{equation}
\Omega={1\over \sqrt{N}}
\left(1-{1\over 8}+\ldots +O\left({1\over N}\right)\right)
\end{equation}
Thus, we arrive at the following approximative expression for the order parameter
at $N\gg 1$:
\begin{equation}
\label{LARGE_NOISE_N}
\Omega(\eta=2\pi)\approx {7\over 8} {1\over \sqrt{N}}\,.
\end{equation}

\section*{Appendix C: Integrals for ${\bf M=2}$ and ${\bf M=3}$}
%-------------- New Insert

In order to calculate the angular integral $K_C^1(M)$, see Eq. (\ref{OMEGA_KC1DEF}), and integrals of similar type,
the average angle $\Phi$ is expressed by means of the local order parameter vector
${\bf L}_M=(L_{M,x},L_{M,y})$,
defined as,
\begin{equation}
\label{LOCAL_ORDER1}
{\bf L}_M=\sum_{i=1}^M\,{\bf \hat{n}}_i
\end{equation}
where ${\bf \hat{n}}_i=(\cos{\alpha_i},\sin{\alpha_i})={\bf v}_i/v_0$ is the normalized velocity vector for agent $i$.
%, and $v$
%is the constant speed of a bird.
The sine and the cosine of the average angle follow as
$\cos{\Phi}=L_x/|L|$ and $\sin{\Phi}=L_y/|L|$.
The average angle is given by $\Phi={\rm atan}(L_y/L_x)$. 
For $M=2$, 
trigonometric addition rules can be used to simplify the integrations,
\begin{eqnarray}
\nonumber
L_x&=&\cos{\alpha_1}+\cos{\alpha_2}=2\cos{\alpha_1+\alpha_2\over 2}\cos{\alpha_1-\alpha_2\over 2} \\
L_y&=&\sin{\alpha_1}+\sin{\alpha_2}=2\sin{\alpha_1+\alpha_2\over 2}\cos{\alpha_1-\alpha_2\over 2} 
\end{eqnarray}
yielding 
\begin{eqnarray}
\nonumber
\Phi&=&{\alpha_1+\alpha_2\over 2}\;\;\;{\rm for}\;\;|\alpha_1-\alpha_2|<\pi \\
\Phi&=&{\alpha_1+\alpha_2\over 2}+\pi\;\;\;{\rm for}\;\;|\alpha_1-\alpha_2|>\pi 
\end{eqnarray}
for $0\leq \alpha_i\leq 2 \pi$.
The integral over $\alpha_1$ and $\alpha_2$ is split into four parts,
\begin{eqnarray}
\nonumber
& &\int_0^{2 \pi}\,d\alpha_1\int_0^{2\pi}\,d\alpha_2\ldots \\
\nonumber
&=&\int_0^{\pi}\,d\alpha_1
\left(\int_0^{\alpha_1+\pi}\,d\alpha_2\ldots+
      \int_{\alpha_1+\pi}^{2 \pi}\,d\alpha_2 \ldots\right) \\
&+&\int_{\pi}^{2\pi}\,d\alpha_1
\left(\int_{\alpha_1-\pi}^{2 \pi}\,d\alpha_2\ldots+
      \int_{0}^{\alpha_1-\pi}\,d\alpha_2\ldots \right)
\end{eqnarray}
where in the first and third part 
$|\alpha_1-\alpha_2|<\pi$, and in the second and fourth term one has
$|\alpha_1-\alpha_2|>\pi$.
All functions under the integral are now products of sine and cosine with a linear combination $a\alpha_1+b\alpha_2$.
Therefore, all integrals of the kind shown in Eq. (\ref{OMEGA_KC1DEF}) can be avaluated analytically for $M=2$.
For example, one finds $K_C^1(M=2)=1/\pi$, see Table I.
More details and information about how to exactly evaluate collision integrals for $M=3$ and $M\rightarrow \infty$ will be given 
elsewhere \cite{chou_12}.
%
%-------------- End Insert

In order to perform the order parameter calculations and the linear stability analysis
for systems with three-body interactions, $M=3$, the following integrals are needed:
\begin{eqnarray}
K_{mpqr}^{cccc}&\equiv & 
\langle
\cos(m\Phi)
\cos(p\tilde{\theta}_1)\cos(q\tilde{\theta}_2)\cos(r\tilde{\theta}_3)
\rangle
\nonumber\\
K_{mpqr}^{sccs}&\equiv & 
\langle
\sin(m\Phi)
\cos(p\tilde{\theta}_1)\cos(q\tilde{\theta}_2)\sin(r\tilde{\theta}_3)
\rangle
\nonumber\\
K_{mpqr}^{scsc}&\equiv & 
\langle
\sin(m\Phi)
\cos(p\tilde{\theta}_1)\sin(q\tilde{\theta}_2)\cos(r\tilde{\theta}_3)
\rangle
\nonumber\\
K_{mpqr}^{sscc}&\equiv & 
\langle
\sin(m\Phi)
\sin(p\tilde{\theta}_1)\cos(q\tilde{\theta}_2)\cos(r\tilde{\theta}_3)
\rangle
\end{eqnarray}
with $\langle\ldots\rangle\equiv\int_0^{2\pi} d\tilde{\theta}_1\int_0^{2\pi}d\tilde{\theta}_2
\int_0^{2\pi} d\tilde{\theta}_3/(2\pi)^3$.
Using the definitions
\begin{eqnarray}
K_{mpqr}^{cccc}&\equiv & S_1 K
\nonumber\\ 
K_{mpqr}^{sccs}&\equiv & S_2 K
\nonumber\\
K_{mpqr}^{scsc}&\equiv & S_3 K
\nonumber\\
K_{mpqr}^{sscc}&\equiv & S_4 K,
\end{eqnarray}
where $K$ and $S_i$ depend on the quadrupel $(m,p,q,r)$,
the numerical values of these integrals for $m\le 5$ and $p+q+r\le 6$ can be constructed from Table II. %\ref{TAB_M3_INT}.
\begin{table}[h]
\label{TAB_M3_INT}
\[\begin{array}{cccclrrrr}
 m & p & q & r & K & S_1 & S_2 & S_3 & S_4 \\
\hline
 0 & 0 & 0 & 0 & 1 & 1 & 0 & 0 & 0 \\
 1 & 0 & 0 & 1 & 0.262433 & 1 & 1 & 0 & 0 \\
 1 & 0 & 1 & 2 & 0.012774 & -1 & -1 & 1 & 0 \\
 1 & 0 & 2 & 3 & 0.005155 & -1 & -1 & 1 & 0 \\
%1 & 0 & 3 & 4 & 0.004910 & 1 & 1 & -1 & 0 \\
 1 & 1 & 1 & 1 & 0.023223 & -3 & -1 & -1 & -1 \\
 1 & 1 & 1 & 3 & 0.005575 & 1 & 1 & -1 & -1 \\
 1 & 1 & 2 & 2 & 0.016024 & 1 & 0 & 0 & 1 \\
%1 & 1 & 2 & 4 & 0.001425 & -1 & -1 & 1 & 1 \\
%1 & 1 & 3 & 3 & 0.001433 & -1 & 0 & 0 & -1 \\
%1 & 2 & 2 & 3 & 0.005723 & -1 & 1 & -1 & -1 \\
 2 & 0 & 0 & 2 & 0.108998 & 1 & 1 & 0 & 0 \\
 2 & 0 & 1 & 1 & 0.097751 & 1 & 1 & 1 & 0 \\
 2 & 0 & 1 & 3 & 0.005624 & -1 & -1 & 1 & 0 \\
 2 & 0 & 2 & 4 & 0.008794 & -1 & -1 & 1 & 0 \\
 2 & 1 & 1 & 2 & 0.048875 & -1 & -1 & 0 & 0 \\
 2 & 1 & 1 & 4 & 0.005624 & 1 & 1 & -1 & -1 \\
 2 & 1 & 2 & 3 & 0.007209 & 1 & 1 & -1 & 1 \\
 2 & 2 & 2 & 2 & 0.017229 & 3 & 1 & 1 & 1 \\
 3 & 0 & 0 & 3 & 0.058045 & 1 & 1 & 0 & 0 \\
 3 & 0 & 1 & 2 & 0.038323 & 1 & 1 & 1 & 0 \\
 3 & 0 & 1 & 4 & 0.002354 & -1 & -1 & 1 & 0 \\
%3 & 0 & 2 & 5 & 0.009288 & -1 & -1 & 1 & 0 \\
 3 & 1 & 1 & 1 & 0.069670 & 1 & 1 & 1 & 1 \\
 3 & 1 & 1 & 3 & 0.033448 & -1 & -1 & 0 & 0 \\
%3 & 1 & 1 & 5 & 0.005079 & 1 & 1 & -1 & -1 \\
 3 & 1 & 2 & 2 & 0.024036 & -1 & -1 & -1 & 1 \\
%3 & 1 & 2 & 4 & 0.004274 & 1 & 1 & -1 & 1 \\
%3 & 2 & 2 & 3 & 0.034337 & 1 & 1 & 0 & 0 \\
 4 & 0 & 0 & 4 & 0.041514 & 1 & 1 & 0 & 0 \\
 4 & 0 & 1 & 3 & 0.011247 & 1 & 1 & 1 & 0 \\
 4 & 0 & 1 & 5 & 0.001433 & -1 & -1 & 1 & 0 \\
 4 & 0 & 2 & 2 & 0.005624 & -1 & -1 & -1 & 0 \\
 4 & 1 & 1 & 2 & 0.048875 & 1 & 1 & 1 & 1 \\
 4 & 1 & 1 & 4 & 0.022494 & -1 & -1 & 0 & 0 \\
 4 & 1 & 2 & 3 & 0.014417 & -1 & -1 & -1 & 1 \\
 5 & 0 & 0 & 5 & 0.032937 & 1 & 1 & 0 & 0 \\
 5 & 0 & 1 & 4 & 0.003923 & 1 & 1 & 1 & 0 \\
%5 & 0 & 1 & 6 & 0.001020 & -1 & -1 & 1 & 0 \\
 5 & 0 & 2 & 3 & 0.025773 & -1 & -1 & -1 & 0 \\
 5 & 1 & 1 & 3 & 0.027873 & 1 & 1 & 1 & 1 \\
%5 & 1 & 1 & 5 & 0.016931 & -1 & -1 & 0 & 0 \\
 5 & 1 & 2 & 2 & 0.040060 & 1 & 1 & 1 & 1 \\
%5 & 1 & 2 & 4 & 0.007123 & -1 & -1 & -1 & 1 \\
%5 & 1 & 3 & 3 & 0.003582 & -1 & -1 & -1 & 1 \\
\end{array}\]
\caption{Integrals for $M=3$ defined in Appendix C.}
\end{table}
%%%%%%%%%%%
%%%%%%%%%%%
\section*{Appendix D: Effect of collisional momentum transfer}
Assume spatial variations with wavelengths $2\pi/k$ that are much larger than the effective collision range $R_\text{eff}$. %\sim 1\sqrt{\rho}$.
In this limit, all fields including the distribution function $f$ and the density $\rho$ are constant inside a circle which is centered 
around position ${\bf x}$ and has radius $R_\text{eff}$.
The exponential prefactor in Eq. (\ref{COLLISION4}) becomes small for radii $R_j>R_\text{eff}$ and thus very effectively 
suppresses
errors when $f$ and $\rho$ are crudely approximated
far away from ${\bf x}$. This allows us to approximate the value of the density in the integral of Eq. (\ref{LOCAL_M})
by $\rho({\bf x})$ for any radius $R_j$. Then the integrand is constant and the simple result $\overline{M}_j=\pi R_{j+1}^2 \rho({\bf x})$ is obtained.
% for
%\Delta R\rightarrow 0$.
Similarly, we can formally replace the distribution functions 
$f(\tilde{\theta}_i,\mathbf{x}_i)$ by their value at the point ${\bf x}$, $f(\tilde{\theta}_i,\mathbf{x})$, where $i=2, 3, \cdots, M$, in 
the collision integral (\ref{COLLISION4}).
This effectively ignores field variations within typical collision distances. 
After integrating over the positions of all the collision partners, the equation becomes,
%The variation of distribution function inside the collision circle can be ignored and the distribution function $f(\tilde{\theta}_i,\mathbf{x}_i)$ may be replaced by $f(\tilde{\theta}_i,\mathbf{x})$, where $i=2, 3, \cdots, M$, in the collision integral (\ref{COLLISION4}). After integrating over the positions of all the collision partners, the Enskog equation becomes,
\begin{eqnarray}
&&f(\theta,\mathbf{x}+\tau \mathbf{v}, t+\tau)
\nonumber\\
&&~=\frac{1}{\rho(\mathbf{x})^{M-1}}
\int_{-\eta/2}^{\eta/2}\frac{d\xi}{\eta}
\int d\tilde{\theta}_1 d\tilde{\theta}_2 \ldots d\tilde{\theta}_M 
\hat{\delta}(\theta-\xi-\Phi_1)
\nonumber\\
&&~f(\tilde{\theta}_1,\mathbf{x})f(\tilde{\theta}_2,\mathbf{x})\ldots
f(\tilde{\theta}_M,\mathbf{x}).
\label{Enskog_new}
\end{eqnarray}
The approximative collision term on the r.h.s. contains only information from the point ${\bf x}$ but not from surrounding points anymore.
In an Enskog equation, it is the differences in the field values around point ${\bf x}$, which account for the so-called collisional momentum
transfer. Therefore, we have effectively removed this transfer, and Eq. ({\ref{Enskog_new}) can be seen as the Boltzmann limit of the more general
Enskog-like kinetic equation, Eq. (\ref{COLLISION4}). 
Moreover, for $M=2$, Eq. (\ref{Enskog_new}) can be directly compared with the equation postulated by Peshkov et al. \cite{peshkov_12}.
%, where 
%they have a Gaussian distributed noise instead of an uniform one and an infinitesimal time step.
%
By rewriting Eq. (\ref{Enskog_new}) into the format of Eq. (\ref{eq.Enskog_MF_2}) in order to investigate the linear stability, 
one sees that the original integrand $\mathcal{F}(\tilde{\theta}_1,\tilde{\theta}_2,\ldots\tilde{\theta}_M)$ is replaced by the following,
\begin{eqnarray}
&&\mathcal{F}'(\tilde{\theta}_1,\tilde{\theta}_2,\ldots\tilde{\theta}_M)
\\ 
&&=\left(\prod_{i=1}^M\bar{f}_i\right)  \left[1- (M-1)\frac{\delta g_0}{g_0}\right]
+ M\delta{f}_1 \prod_{i=2}^M\bar{f}_i.
\nonumber
\label{eq.Enskog_MF_2_kernel_3}
\end{eqnarray}
Again, nonlinear perturbations of order  $\mathcal{O}(\delta^2)$ were dropped in this expansion.
Comparing this with Eq. (\ref{eq.Enskog_MF_2_kernel_2}), we note that it can be formally obtained by setting 
$z=0$ in the original kernel.
%the only difference is that all the $z$ dependence disappears 
%in the approximated integral kernel, which means $z=0$ since ${}_1\mathrm{F}_1(a,b,0)=1$. 
Using the dimensionless wavenumber $\hat{k}\equiv \lambda k$ and the definition of effective collision radius, Eq. (\ref{REFF}), 
we have $z=-R_{\mathrm{eff}}^2\hat{k}^2/(4M\lambda^2)$. 
Thus, for $z$ to approach zero, either $\Lambda = \lambda/R_\mathrm{eff}$ must go to infinity or the wave number must be zero. 
This confirms that the approximative Boltzmann-like equation (\ref{Enskog_new}) and hence the kinetic equation proposed in Ref. \cite{peshkov_12}
present the zero wave number or infinite $\Lambda$ limit 
where collisional momentum transfer is irrelevant. 
For $M=2$ we compare the stability of the Enskog and the approximative Boltzmann equations in Fig. \ref{FIG17} where the 
growth rates for perturbations of the ordered state are plotted. 
As expected we observe that both approaches agree exactly for zero $k$ and very large $\Lambda\ge 8$.
When $\Lambda$ is decreased, collisional momentum transfer becomes more important and the growth rates become more negative
compared to the transfer-free case. Especially for $\Lambda=1/4$ the difference is very pronounced, even at small $\lambda k<1$. 
Hence, collisional momentum transfer makes the ordered phase even more stable.
This is consistent with results for other systems \cite{ihle_05}, where collisional momentum transfer leads to an additional contribution to
the viscosity and thermal conductivity, causing a stronger attenuation of sound and shear modes. 
In Fig. \ref{FIG17} one also sees that the sound mode is not as strongly affected by collisional momentum transfer as the other modes.
\begin{figure}
\begin{center}
\vspace{0cm}
\includegraphics[width=3.1in,angle=0]{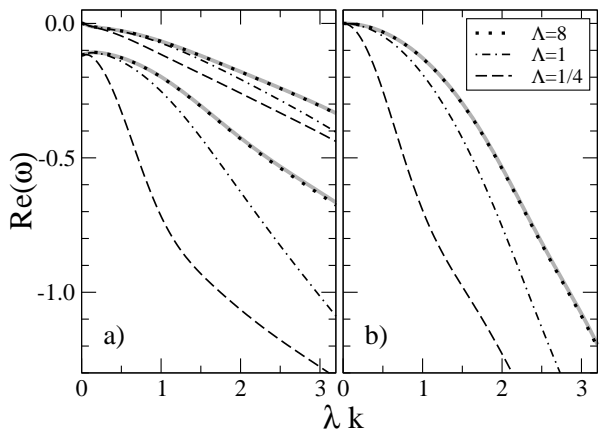}
\caption{Real part of the growth rate $\omega$ as a function of wave number for $M=2$,  at $\eta=0.99\eta_C$ and with density $\rho_0=0.1$.
Part a) shows the first two longitudinal modes, whereas the first transversal mode is shown in b).
The results of the Enskog-like equation, Eq. (\ref{eq.Enskog_MF_2}), for various ratios of mean free path to effective radius, 
$\Lambda$, are shown in dotted, dot-dashed, and dashed lines, whereas
the results for the Boltzmann approach, Eq. (\ref{Enskog_new}) are given by gray solid lines. 
For the case where $\Lambda$ is as large as $8$, the curves of the original Enskog-like equation collapse with those of the Boltzmann approximation.
}
\label{FIG17}
\end{center}
\vspace*{-0.5ex}
\end{figure}
%%%%%%%%%%%
%%%%%%%%%%%

\end{document}